\documentclass[letterpaper,aps,prd,twocolumn,tightenlines,preprintnumbers,nofootinbib,showkeys,superscriptaddress,longbibliography]{revtex4-1}
\pdfoutput=1

\usepackage[dvipsnames]{xcolor}
\usepackage{graphicx,epic,eepic,epsfig}
\usepackage[export]{adjustbox}
\usepackage{tikz,contour} 
\usetikzlibrary{shapes,arrows,positioning,automata,backgrounds,calc,er,patterns}
\usepackage[compat=1.1.0]{tikz-feynman}

\usepackage{fullpage}
\usepackage{relsize}
\usepackage[left=1.65cm,right=1.65cm,top=1.6cm,bottom=1.65cm]{geometry}

\usepackage{XCharter}
\usepackage{mathptmx}
\usepackage[T1]{fontenc}
\usepackage{pifont}
\usepackage{amsfonts,amsmath,amssymb,amsbsy,bm,mathtools,slashed,latexsym,esvect,simpler-wick}

\usepackage{float,multirow,array,cancel,dcolumn}
\usepackage{enumitem}
\usepackage{hyperref}
\hypersetup{colorlinks=true,citecolor=red,linkcolor=NavyBlue,urlcolor=NavyBlue}
\usepackage[caption=false,labelformat=simple]{subfig}
\usepackage{lipsum}

\usepackage{comment}
\usepackage{shorthand}
\usepackage{natbib}

\newcommand{\cmark}{\ding{51}}%
\newcommand{\xmark}{\ding{55}}%

\begin{document}
\relscale{1.05}

\title{Vector leptoquark contributions to lepton dipole moments}

\author{Arvind Bhaskar}
\email{arvind.bhaskar@iopb.res.in}
\affiliation{Institute of Physics, Sachivalaya Marg, Bhubaneswar 751 005, India}

\author{Diganta Das}
\email{diganta.das@iiit.ac.in}
\affiliation{Center for Computational Natural Sciences and Bioinformatics, International Institute of Information Technology, Hyderabad 500 032, India}

\author{Soumyadip Kundu}
\email{soumyadip23@iisertvm.ac.in}
\affiliation{Indian Institute of Science Education and Research Thiruvananthapuram, Vithura, Kerala, 695 551, India}

\author{Anirudhan A. Madathil}
\email{anirudhan.alanthatta@utah.edu}
\affiliation{Department of Physics and Astronomy, University of Utah, Salt Lake City, UT 84112, USA}

\author{Tanumoy Mandal}
\email{tanumoy@iisertvm.ac.in}
\affiliation{Indian Institute of Science Education and Research Thiruvananthapuram, Vithura, Kerala, 695 551, India}

\author{Subhadip Mitra}
\email{subhadip.mitra@iiit.ac.in}
\affiliation{Center for Computational Natural Sciences and Bioinformatics, International Institute of Information Technology, Hyderabad 500 032, India}
\affiliation{Center for Quantum Science and Technology, International Institute of Information Technology, Hyderabad 500 032, India}

\begin{abstract}
\noindent
Leptoquarks (LQs) can contribute to the magnetic and electric dipole moments of charged leptons, which the current experiments have measured with good accuracy. We revisit the parameter spaces of TeV-scale vector LQs that contribute to these observables and study how these models fare against the LHC bounds. We show that significant portions of the parameter space are excluded when the current LHC data is utilised effectively. We find that only $U_1$ and $V_2$ can explain the observed positive shift in $(a_\mu^{\rm exp} - a_\mu^{\rm SM})$ through a lepton chirality-flipping contribution with $\mathcal O(1)$ LQ-quark-lepton coupling. We also see how these two LQs can fit the electron dipole moment and atomic parity violation measurements. We find that the current electric dipole moment measurements of the muon cannot restrain the LQ-quark-lepton couplings within perturbative regions. 
\end{abstract}

\maketitle 
\section{Introduction}

\noindent
The Standard Model (SM) predictions for the lepton anomalous magnetic dipole moments [AMDMs, $(g-2)_{\ell}$] are precise, and the light-lepton ones are also measured experimentally with remarkable accuracy (see, e.g.,~\cite{ParticleDataGroup:2024cfk}). Therefore, even a minor discrepancy between the theoretical and experimental values poses severe challenges to the two sides. Robust and consistent experimental measurements prompt a revisit to the theoretical computations, which are highly complex. Conversely, a strong confidence in our SM calculations demands a re-evaluation of the experiments and more precise measurements. A disagreement between theoretical and experimental values of $(g-2)_\mu$ has remained unresolved for more than two decades: the SM expectation for the muon AMDM is $a_\mu^{\rm SM}= (g-2)^{\rm SM}_\mu/2 = 116591810(43)\times 10^{-11}$~\cite{Aoyama:2020ynm,*Aoyama:2012wk,*Aoyama:2019ryr,*Czarnecki:2002nt,*Gnendiger:2013pva,*Melnikov:2003xd,*Masjuan:2017tvw,*Colangelo:2017fiz,*Hoferichter:2018kwz,*Gerardin:2019vio,*Bijnens:2019ghy,*Colangelo:2019uex,*Blum:2019ugy,*Colangelo:2014qya,Davier:2017zfy,*Keshavarzi:2018mgv,*Colangelo:2018mtw,*Hoferichter:2019mqg,*Davier:2019can,*Keshavarzi:2019abf,*Kurz:2014wya}, whereas the current experimental world average reported by the Muon $g-2$ Collaboration is $a_\mu^{\rm exp} = 116592059(22)\times 10^{-11}$~\cite{ParticleDataGroup:2024cfk,Muong-2:2023cdq,*Muong-2:2006rrc,*Muong-2:2021ojo}. The experimental value is about $5\sigma$ away from the SM expectation, i.e., $\Delta a_\mu = a_\mu^{\rm exp} - a_\mu^{\rm SM} = (2.49 \pm 0.48)\times 10^{-9}$. If we trust both the SM predictions and experimental measurements, we can take this discrepancy as a hint of the presence of new physics (NP) beyond the SM.

The $(g-2)_\mu$ anomaly has several possible NP explanations (see Ref.~\cite{Athron:2021iuf} for a recent review and the references). Prominent among these are the models featuring leptoquarks (LQs or $\ell_q$)---hypothetical bosons that couple simultaneously to a lepton and a quark. The lepton and quark sectors in the SM are similar---both have three generations---and this parity is necessary to cancel the gauge anomalies. However, these similarities perhaps also suggest a deeper connection between these two sectors. Within the unified theory framework, leptons and quarks naturally couple through LQs~\cite{Georgi:1974sy,Pati:1974yy}.  In various theories with extended gauge symmetries, vector LQs (vLQs) arise when the symmetries break spontaneously. Though less explored than their scalar counterparts (e.g.,  Ref.~\cite{Bigaran:2020jil,*Dorsner:2020aaz,*Bigaran:2021kmn}), vLQ models have been considered as possible solutions to the $(g-2)_\mu$ anomaly~\cite{Queiroz:2014zfa,*Crivellin:2018qmi,*Kowalska:2018ulj,*Datta:2019bzu,*Crivellin:2021rbq,*Du:2021zkq,*Ban:2021tos,*Yu:2021suw,*Cheung:2022zsb,Altmannshofer:2020ywf}. The difficulty in obtaining consistent ultraviolet-complete theories with vLQs has been the main obstacle in this direction. However, some such solutions are known, e.g., the $4321$ model~\cite{DiLuzio:2017vat,*DiLuzio:2018zxy}. In such a framework, one can consistently calculate higher-loop effects involving vLQs. These effects in various vLQ models have been discussed in Ref.~\cite{Fuentes-Martin:2019ign,*Fuentes-Martin:2020luw,*Fuentes-Martin:2020hvc,*Haisch:2022afh}. TeV-scale vLQs that appear in different phenomenological variants of the Pati-Salam 
model~\cite{Assad:2017iib,DiLuzio:2017vat,Calibbi:2017qbu,*Bordone:2017bld,*Barbieri:2017tuq,*Blanke:2018sro,*Greljo:2018tuh,*Bordone:2018nbg,DiLuzio:2018zxy,*Heeck:2018ntp,*Fornal:2018dqn,*Balaji:2018zna,*Balaji:2019kwe,*Iguro:2021kdw,*Gedeonova:2022iac,*FernandezNavarro:2022gst,Baker:2019sli} or the composite GUT framework~\cite{DaRold:2019fiw} can explain various flavour anomalies. In some bottom-up studies, vLQ models with various flavour ansatzes for the new couplings (i.e., vLQ-quark-lepton couplings) have been used to address multiple anomalies simultaneously~\cite{Alonso:2015sja,*Greljo:2015mma,*Fajfer:2015ycq,*Barbieri:2015yvd,*Becirevic:2016oho,*Sahoo:2016pet,*Faroughy:2016osc,*Bhattacharya:2016mcc,*Duraisamy:2016gsd,*Kumar:2018kmr,*Crivellin:2018yvo,*Angelescu:2018tyl,*Chauhan:2018lnq,*Hati:2019ufv,*Cornella:2019hct,*Cheung:2020sbq,*BhupalDev:2020zcy,*Hati:2020cyn,*Angelescu:2021lln,*Cornella:2021sby,*King:2021jeo,*Belanger:2022kvj,*Barbieri:2022ikw,*Garcia-Duque:2022tti,Buttazzo:2017ixm,Aebischer:2022oqe}.

The current LHC data strongly constrains the parameter spaces of TeV-scale LQ models. The nonresonant LQ production (i.e., $t$-channel LQ exchanges) and its interference with the SM can significantly contribute to the invariant-mass distribution of high-$p_T$ lepton pairs. (This distribution is also influenced by additional contributions from resonant pair and single LQ productions, particularly in the low-mass region. Combining both resonant and nonresonant (including interference) contributions and utilising the high-$p_T$ dilepton search data, Refs.~\cite{Mandal:2018kau,*Aydemir:2019ynb,*Bhaskar:2022vgk,*Aydemir:2022lrq,Bhaskar:2021pml,Bhaskar:2023ftn} showed the LQ parameter space to be highly constrained in general.) The bounds from high-$p_T$ di-muon and di-tau data on vLQ couplings can be seen from Ref.~\cite{Bhaskar:2021pml} and Refs.~\cite{Buttazzo:2017ixm,Baker:2019sli,Aebischer:2022oqe}, respectively. Recasting the direct LQ search data in the dilepton-dijet channels~\cite{Mandal:2015vfa,Bhaskar:2023ftn} also gives strong constraints. The LHC phenomenologies of various LQs have been well-explored---they show good prospects at the LHC (see, e.g., Refs.~\cite{Mandal:2015lca,*Chandak:2019iwj,*Bhaskar:2020gkk,*Bhaskar:2021gsy,*Cheung:2023gwm,*Florez:2023jdb}).

Unlike the scalar LQs, a summary of the status of vLQ models in the light of dipole moment and high-energy measurements in the light-lepton sector is still lacking in the literature. In this paper, we review the current status of the vLQ contributions to the $(g-2)_\ell$ measurements against the latest LHC dilepton data. We look at the parameter regions relevant to the electron and muon AMDM measurements. For the promising vLQs, we also find the bounds from the electric dipole moment (EDM) measurements. The rest of the paper is planned as follows. In the next section, we look at the current experimental status of the lepton dipole moments. In Sec~\ref{sec:vlq}, we discuss the vLQ models; in Sec~\ref{sec:gm2fit}, we see how these models contribute to $(g-2)_\ell$ and show the results of our parameter scans in Sec.~\ref{sec:results}. We look at the EDM limits in Sec.~\ref{sec:EDM}. We also look at the limits on the first-generation fermion-vLQ couplings from the measurement of atomic parity violation (APV) in the $\prescript{133}{55}{\mathbf{Cs}}$ nucleus in Sec~\ref{sec:apv} before concluding in Sec.~\ref{sec:conclu}.

\begin{figure}[]
\begin{center}
\subfloat{
\begin{tikzpicture}[thick]
\begin{feynman}
\vertex (i1){\(\mu\)};
\vertex [right= of i1,purple,dot](a){};
\vertex [right= of a](c);
\vertex [right= of c,purple,dot](b){};
\vertex [above = of c](g1);
\vertex  [above = of g1](r){\(\gamma\)};
\vertex [right = of b](f2){\(\mu\)};
\diagram*{
(i1) -- [fermion,momentum=\(p_{1}\)](a)--[edge label'=\(q\)](c)--[fermion,momentum'=\(k\)](b)--[fermion,momentum=\(p_{2}\)](f2),
(a)--[boson,edge label'=\textcolor{purple}{\(V_{LQ}\)},momentum=\textcolor{black}{\(p_{1}-k\)},quarter left,purple](g1)--[boson,edge label'=\textcolor{purple}{\(V_{LQ}\)},momentum=\textcolor{black}{\(p_{2}-k\)},quarter left,purple](b),
(r)--[boson,momentum=\(p_{2}-p_{1}\)](g1),
};
\end{feynman}
\end{tikzpicture}\label{fig:fdB}
}\\
\subfloat{
    \begin{tikzpicture}[thick]
\begin{feynman}
\vertex (i1){\(\mu\)};
\vertex [right= of i1,purple,dot](a){};
\vertex [right= of a](c);
\vertex [right= of c,purple,dot](b){};
\vertex [above = of c](g1){\textcolor{purple}{\(V_{LQ}\)}};
\vertex  [below = of c](r){\(\gamma\)};
\vertex [right = of b](f2){\(\mu\)};
\diagram*{
(i1) -- [fermion,momentum=\(p_{1}\)](a)--[fermion,momentum=\(p_{1}-k\)](c)--[fermion,momentum=\(p_{2}-k\),edge label'=\(q\)](b)--[fermion,momentum=\(p_{2}\)](f2),
(a)--[boson,momentum=\textcolor{black}{\(k\)},quarter left,purple](g1)--[boson,quarter left,purple](b),
(r)--[boson,momentum=\(p_{2}-p_{1}\)](c),
};
\end{feynman}
\end{tikzpicture}\label{fig:fdA}
}
\end{center}
\caption{Feynman diagrams involving the vLQ contribution to the muon $g-2$. \label{fig:FDAMDMVLQ}}
\end{figure}
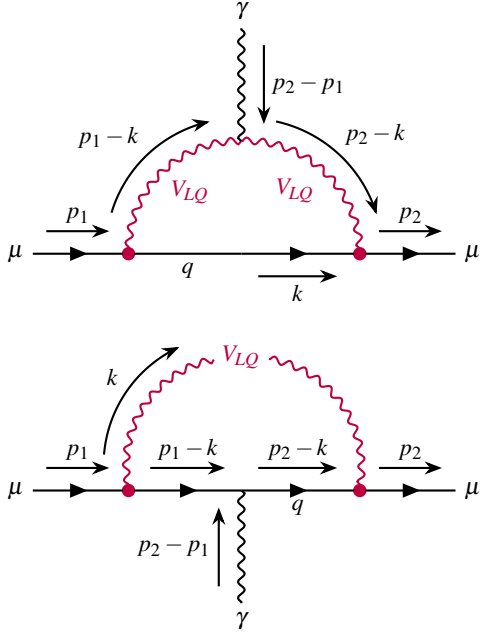

\section{Lepton dipole moments: Current status}
\label{sec:dipmomlep}
\noindent
A lepton of charge $Q_\ell$, mass $m_\ell$, and spin $\bm{S}$ has a spin magnetic moment, $\bm{\mu} = g_\ell (Q_\ell/2m_\ell)\bm{S}$, where $g_\ell$ is the gyromagnetic ratio. Quantum electrodynamics predicts $g_\ell=2$ at the leading order (LO). However, $g_\ell$ receives higher order corrections and shifts from the LO value. The AMDM parametrises this shift as, $a_\ell = (g_\ell-2)/2$. A general $\gamma\ell\ell$ vertex  
can be expressed as,
\begin{align}
\label{eq:gmllvertex}
\Gamma_{\mu}(q)=&\ \gamma_{\mu}F_{1}(q^{2})+i\frac{\sigma^{\mu\nu}q_{\nu}}{2m_\ell}F_{2}(q^{2})+\frac{\sigma^{\mu\nu}\gamma_5q_{\nu}}{2m_\ell}\widetilde{F_{2}}(q^{2})\nn \\
&\ +\dfrac{q^2\gm_\mu-\slashed{q}q_\mu}{m_\ell^2}\gm_5 F_3(q^2)
\end{align}
where $q$ is the momentum exchanged and the functions $F(q^2)$ are the form factors. In the limit $q^{2} \to 0$, $F_1(0) = 1$, $F_2(0)=a_\ell$ and $\widetilde{F_2}(0)= - 2 m_\ell d_\ell/e$, with $d_\ell$ being the EDM. The $F_3(0)$ is the anapole moment, which is not an independent and directly observable quantity for a lepton~\cite{Musolf:1990sa}.  

As mentioned in the introduction, the Muon $g-2$ Collaboration puts the new experimental world average of $a_\mu$ at $a_\mu^{\rm exp} = 116592059(22)\times 10^{-11}$~\cite{Muong-2:2023cdq}, roughly $\sim 5\sigma$ away from the SM expectation, $a_\mu^{\rm SM} = 116591810(43)\times 10^{-11}$~\cite{Aoyama:2020ynm}, i.e., $\Delta a_\mu = a_\mu^{\rm exp} - a_\mu^{\rm SM} = (2.49 \pm 0.48)\times 10^{-9}$. The dominant uncertainty in the calculation comes from the contribution of hadronic vacuum polarization, $a_\mu^{\rm HVP}$. The Muon $g-2$ Theory Initiative adopts a data-driven estimation of $a_\mu^{\rm HVP}$ based on the dispersion relations and $e^+e^-\to~hadrons$ cross sections (see Ref.~\cite{Davier:2017zfy,Keshavarzi:2018mgv,Colangelo:2018mtw,Hoferichter:2019mqg,Davier:2019can,Keshavarzi:2019abf,Kurz:2014wya} for this ongoing endeavour). However, a first-principle lattice-QCD calculation of $a_\mu^{\rm HVP}$ by the BMW Collaboration~\cite{Borsanyi:2020mff} deviates from the data-driven calculation by about $2.1\sigma$, leading to an overall discrepancy of only $1.6\sigma$ with the experimental results. The recent CMD-3 result also reduces the tension between the experimental value of the $a_\mu$ and the SM prediction~\cite{CMD-3:2023alj}. 

Electron AMDM measurements also show some discrepancies, although not as significant as the muon. The $\prescript{133}{55}{\mathbf{Cs}}$-based experiment yielded a $2.4\sg$ negative shift in $\Dl a_e$:  
$\Delta a_e^{Cs} = a_e^{\rm exp} - a_e^{\rm SM} = (-8.8 \pm 3.6)\times 10^{-13}$~\cite{Parker:2018vye}, whereas the $\prescript{87}{37}{\mathbf{Rb}}$-based experiment yielded a positive shift of $1.6\sg$: $\Delta a_e^{Rb} = a_e^{\rm exp} - a_e^{\rm SM} = (4.8 \pm 3.0)\times 10^{-13}$~\cite{Morel:2020dww}. Note that the central values of these two measurements have opposite signs with large error bars. 

As the $\tau$ lepton is a short-lived particle, measuring the $\tau$ AMDM is challenging. Recently, the CMS collaboration has obtained the most precise limit on $a_\tau$ at $0.0009^{+0.0032}_{-0.0031}$~\cite{CMS:2024skm}. The previous bound ($95$\% CL) by the DELPHI collaboration was much weaker: $-0.052< a_\tau < 0.013$~\cite{DELPHI:2003nah,ParticleDataGroup:2022pth}. The theoretical calculations within the SM give a value, $a^{SM}_\tau = 117721(5) \times 10^{-8}$~\cite{Eidelman:2007sb}.

In principle, measuring the EDMs of charged leptons could also be another way to look for NP. For all three charged leptons, the current experimental upper bounds on their EDMs are several orders higher than their SM values~\cite{Mahanta:1996er,Yamaguchi:2020eub,Yamaguchi:2020dsy}---leaving a significant space for NP contributions. The current upper bounds are: $|d_e| < 1.1\times 10^{-29}$~e~cm (90\% CL)~\cite{ACME:2018yjb} for the electron, $|d_\mu| < 1.8\times 10^{-19}$~e~cm (95\% CL)~\cite{Muong-2:2008ebm} for the muon, and $-2.2 < \textrm{Re}(d_\tau) < 4.5$~($10^{-17}$~e~cm) and $-2.5 < \textrm{Re}(d_\tau) < 0.8$~($10^{-17}$~e~cm)~\cite{Belle:2002nla} for the tau. The recent limit on $d_\tau$ by CMS collaboration is $|d_\tau| < 2.9\times 10^{-17}$~e~cm ($95$\% CL)~\cite{CMS:2024skm}. 


\section{Lepton AMDM and vLQ models}
\label{sec:vlq}

\noindent
A vLQ can be singlet, doublet, or triplet under the weak interaction. We summarise the representations and the interactions of the vLQs that can couple to a quark and a charged lepton~\cite{Buchmuller:1986zs,Dorsner:2016wpm} in Table~\ref{tab:vLQYuk}. The couplings $(x_n^{XX})_{ij}$ or $(\widetilde x_n^{XX})_{ij}$ (where $n=\left\{1,\,2,\,3\right\}$ indicates the vLQ representation and $XX=\left\{LL,\,RR,\,RL,\,LR\right\}$ shows the chiralities of the quark and the lepton) are $3\times 3$ matrices in the generation space.  
These are, in general, complex and can act as a source of $CP$ violation. A vLQ with a tree-level coupling to the muon and a quark contributes to $(g-2)_\mu$ at the one-loop level; Fig.~\ref{fig:FDAMDMVLQ} shows the representative Feynman diagrams. LQs without a tree-level $q\mu\ell_q$ coupling can still contribute to $(g-2)_\mu$ at higher-loop levels. In that case, however, the coupling(s) relevant to the $(g-2)_\mu$ anomaly must be large enough to compensate for the extra loop suppression. Here, we only focus on those vLQs that contribute to $(g-2)_\ell$ at the one-loop level.

In principle, vLQs can have additional diquark interactions leading to proton instability. However,  it is possible to construct ultraviolet-complete models with TeV-scale vLQs evading proton lifetime bounds~\cite{Assad:2017iib}. In such models, the baryon number-violating operators are suppressed or forbidden by some discrete symmetry. For phenomenological purposes, we ignore the diquark couplings. We also neglect the possibility of vLQs interacting with another new particle. We consider up-aligned and down-aligned scenarios (where the vLQ mass basis aligns with the mass basis of the up-type or down-type quarks, respectively) for the vLQs coupling to the left-handed quarks separately. There is only one scenario for the vLQs coupling exclusively to right-handed quarks.

\subsection{Weak-singlet vector leptoquarks}
\noindent 
There are two weak-singlet vLQs---$U_1(\mathbf{3},\mathbf{1},2/3)$ and $\widetilde{U}_1(\mathbf{3},\mathbf{1},5/3)$---which can contribute to $(g-2)_\ell$ in principle. Their gauge interactions and mass terms can be generically written as
\begin{align}\label{singlet-Yukawa-gauge}
\mathcal{L} \supset& -\dfrac{1}{2} V_{\mu\nu}^\dagger V^{\mu\nu}+i g_s V_\mu^\dagger T_a V_\nu \left(\kappa_s G^{\mu\nu}_a + \widetilde \kappa_s \widetilde G^{\mu\nu}_a \right) \nn\\
&\ + i g^\prime Y V_\mu^\dagger V_\nu \left( \kappa_Y B^{\mu\nu} + \widetilde \kappa_Y  \widetilde B^{\mu\nu}\right) + M_{V}^{2}V_{\mu}^{\dagger}V^{\mu},
\end{align}
where $V^{\mu\nu} = D^\mu V^\nu - D^\nu V^\mu$ is the  field-strength tensor of the vLQ $V^\m$. Here, $G_a^{\mu\nu}$ and $B^{\mu\nu}$ are the SM gauge field-strength tensors and $\widetilde G^{a\,\mu\nu} = \frac{1}{2} \epsilon^{\mu\nu\rho\sigma} G^a_{\rho\sigma}$ and $\widetilde B^{\mu\nu}= \frac{1}{2} \epsilon^{\mu\nu\rho\sigma} B_{\rho\sigma}$ are the corresponding dual tensors. The anomalous couplings of vLQs are denoted by $\kappa$~\cite{Blumlein:1996qp}.
After the electroweak-symmetry breaking, the electromagnetic interactions of $V^\m$ can be written as 
\begin{align}\label{eq: singlet photon}
\mathcal{L} \supset &\ ieQ\Big\{ (\partial^{\mu} V^{\nu} - \partial^{\nu} V^{\mu})A_{\mu}V^{\dagger}_{\nu} - (\partial_{\mu} V_{\nu}^{\dagger} - \partial_{\nu} V_{\mu}^{\dagger})A^{\mu}V^{\nu}  \nonumber\\
&\ -\kappa_Y(V_{\mu}V_{\nu}^{\dagger}-V_{\nu}V_{\mu}^{\dagger})\partial^{\mu}A^{\nu}\nn\\
&\ -\dfrac{1}{2}\widetilde{\kappa}_Y\epsilon^{\mu\nu\rho\sigma}(V_{\rho}V_{\sigma}^{\dagger}-V_{\sigma}V_{\rho}^{\dagger})\partial_{\mu}A_{\nu}\Big\}.
\end{align}

The interactions of $U_{1}$ and $\widetilde{U}_{1}$ with the SM quarks and leptons are shown in Table~\ref{tab:vLQYuk}. With appropriate flavour ansatzes, the singlet vLQs can produce the required positive shift to $a_\mu$ from the SM value to the experimental value. 
Their contributions to $a_{\m}$ can be obtained from the general expression shown in Appendix~\ref{sec:generalgM2}.
These expressions agree with the ones given in Ref.~\cite{Altmannshofer:2020ywf}.\footnote{  
There, it is also pointed out that $\kappa_s = \kappa_Y = 1$ and $\widetilde \kappa_s = \widetilde \kappa_Y = 0$, in the UV-complete models where the vLQ is the carrier of a spontaneously broken gauge symmetry. Otherwise, these parameters are free in general~\cite{Belyaev:2005ew}. Since the dual terms are $CP$-violating, the nonzero $\widetilde{\kappa}$'s parametrise additional sources of $CP$ violation.} From Eq.~\eqref{eq:LQAMDM}, we see that for free $x_1^{XX}$ couplings, $a_\ell$ has a logarithimic dependence on the UV-cutoff scale $\Lambda$ unless $\kappa_Y=1$ and $\widetilde{\kappa}_Y=0$. [The appearance of the divergent terms is the consequence of adding a gauge boson by hand without considering the full gauge structure. In UV complete theories, $\kappa_Y$ and $\widetilde{\kappa}_Y$ can take different values since the divergences are cancelled by the presence of some other field(s).] The divergent terms vanish for $\kappa_Y=1$ and $\widetilde{\kappa}_Y=0$. Hence, we retain this choice throughout the paper to avoid divergences in our calculations. 

There are two types of contribution to $a_{\mu}$ from a $U_1$---a chirality-flipping part that goes as $x_1^{LL}x_1^{RR}$ and is proportional to $m_\mu m_q$ (where $q$ is the quark that runs in the loops; for $U_1$ it is a down type quark) and a chirality-preserving part proportional to $(|x_1^{LL}|^{2}+|x_1^{RR}|^{2})$ and $m_\m^2$. When $q$ is a bottom quark, the chirality-flipping contribution is more significant for couplings of similar magnitudes. If the quark in the loop is a strange quark, both the chirality-flipping and chirality-preserving terms become comparable as $m_\mu \sim m_s$. 

The $\widetilde{U}_1$ has only right-handed couplings. Therefore, it contributes only through the chirality-preserving terms. For $\kappa_{Y}=1$, the $\widetilde{U}_1$ leads to a positive $\Dl a_\mu$. However, since the chirality-preserving contributions are proportional to $m_\m^2$, one needs relatively larger values of $x^{RR}_1$ to address the $(g-2)_\mu$ anomaly (this is true for vLQs with only left or right couplings in general).

\begin{table*}[hbt]
\caption{List of vLQs and their representations under the SM gauge groups. In the first column, the first and the second arguments within the parentheses indicate the representations under $SU(3)_c$ and $SU(2)_L$, respectively and the third argument is the hypercharge. The electric charge of a vLQ is given by $Q = T_{3} + Y$ where $T_{3}$ is the third component of its isospin and $Y$ is its hypercharge. We present the vLQ interactions in the up- and down-aligned scenarios for all vLQs~\cite{Blumlein:1992ej,Dorsner:2016wpm}.}
\label{tab:vLQYuk}
\centering{\footnotesize\renewcommand\baselinestretch{2}\selectfont
\begin{tabular*}{\textwidth}{l @{\extracolsep{\fill}}cc}
\hline \hline vLQ & Down-aligned vLQ interactions & Up-aligned vLQ interactions \\ \hline\hline
\multirow{2}{*}{$U_1~({\mathbf 3},{\mathbf 1},2/3)$} & $(Vx_{1}^{LL})_{ij}\ \overline{u}_{L}^{i}\gamma^{\mu}\nu^{j}_{L}U_{1,\mu} + (x_1^{LL})_{ij}\ \overline{d}_{L}^{i}\gamma^{\mu}e^{j}_{L}U_{1,\mu}$ & $(x_1^{LL})_{ij}\ \overline{u}_{L}^{i}\gamma^{\mu}\nu^{j}_{L}U_{1,\mu} + (V^{\dagger}x_1^{LL})_{ij}\ \overline{d}_{L}^{i}\gamma_{\mu}e^{j}_{L}U_{1,\mu}$          \\
& $+ (x_1^{RR})_{ij}\ \overline{d}_R^{i}\gamma^{\mu}e^{j}_{R}U_{1,\mu}$ & $+ (x_1^{RR})_{ij}\ \overline{d}_R^{i}\gamma^{\mu}e^{j}_{R}U_{1,\mu}$ \\
$\widetilde{U}_{1}~({\mathbf 3},{\mathbf 1},5/3)$ & \multicolumn{2}{c}{$(\widetilde{x}_1^{RR})_{ij}\ \overline{u_{R}}^{i}\gamma^{\mu}e^{j}_{R}\widetilde{U}_{1}$} \\
\multirow{2}{*}{$V_{2}~(\overline{{\mathbf 3}},{\mathbf 2},{5/6})$} & $-(x_2^{RL})_{ij}\ \overline{d^{C}_{R}}^{i}\gamma^{\mu}\nu^{j}_{L}V^{1/3}_{2,\mu} + (x_2^{RL})_{ij}\ \overline{d^{C}_{R}}^{i}\gamma^{\mu}e^{j}_{L}V^{4/3}_{2,\mu}$ & $-(x_2^{RL})_{ij}\ \overline{d^{C}_{R}}^{i}\gamma^{\mu}\nu^{j}_{L}V^{1/3}_{2,\mu} + (x_2^{RL})_{ij}\ \overline{d^{C}_{R}}^{i}\gamma^{\mu}e^{j}_{L}V^{4/3}_{2,\mu}$ \\
& $+ (V^{*}x_{2}^{LR})_{ij}\ \overline{u^{C}_L}^{i}\gamma^{\mu}e^{j}_{R}V^{1/3}_{2,\mu} - (x_2^{LR})_{ij}\ \overline{d^{C}_L}^{i}\gamma^{\mu}e^{j}_{R}V^{4/3}_{2,\mu}$ & $+ (x_2^{LR})_{ij}\ \overline{u^{C}_L}^{i}\gamma^{\mu}e^{j}_{R}V^{1/3}_{2,\mu} - (V^{\dagger}x_{2}^{LR})_{ij}\ \overline{d^{C}_L}^{i}\gamma^{\mu}e^{j}_{R}V^{4/3}_{2,\mu}$ \\
$\widetilde{V}_{2}~(\overline{{\mathbf 3}},{\mathbf 2},{-1/6})$ & \multicolumn{2}{c}{$-(\widetilde{x}_2^{RL})_{ij}\
\overline{d^{C}_{R}}^{i}\gamma^{\mu}e^{j}_{L}\widetilde{V}^{1/3}_{2} + (\widetilde{x}_2^{RL})_{ij}\
\overline{u^{C}_{R}}^{i}\gamma^{\mu}\nu^{j}_{L}\widetilde{V}^{-2/3}_{2}$} \\
\multirow{2}{*}{$U_{3}~({\mathbf 3}, {\mathbf 3},2/3)$} & $-(x_3^{LL})_{ij}\ \overline{d_{L}}^{i}\gamma^{\mu}e^{j}_{L}U^{2/3}_{\mu} 
+(Vx^{LL}_3)_{ij}\ \overline{u_{L}}^{i}\gamma^{\mu}\nu^{j}_{L}U^{2/3}_{\mu}$ 
& $- (V^{\dagger}x_3^{LL})_{ij}\  \overline{d_{L}}^{i}\gamma^{\mu}e^{j}_{L}U^{2/3}_{\mu}+(x_3^{LL})_{ij}\ \overline{u_{L}}^{i}\gamma^{\mu}\nu^{j}_{L}U^{2/3}_{\mu}$ \\
& $+\sqrt{2}(x_3^{LL})_{ij}\ \overline{d_{L}}^{i}\gamma^{\mu}\nu^{j}_{L}U^{-1/3}_{3}+\sqrt{2}(Vx^{LL}_{3})_{ij}\ \overline{u_{L}}^{i}\gamma^{\mu}e^{j}_{L}U^{5/3}_{\mu}$ & $+\sqrt{2}(V^{\dagger} x_3^{LL})_{ij}\ \overline{d_{L}}^{i}\gamma^{\mu}\nu^{j}_{L}U^{-1/3}_{\mu} + \sqrt{2} (x_3^{LL})_{ij}\ \overline{u_{L}}^{i}\gamma^{\mu}e^{j}_{L}U^{5/3}_{\mu}$\\
\hline
\hline
\end{tabular*}}
\end{table*}

\subsection{Weak-doublet vector leptoquark}
\noindent
For a weak-doublet vLQ, the interaction and mass terms can be written as
\begin{align}\label{eq:doublet-SM}
\mathcal{L} 
  \supset&\  -\frac{1}{2} \textrm{Tr} [\vv{V}_{\mu\nu}^\dagger \vv{V}^{\mu\nu}]+ M_{V}^{2}\vv{V}^{\dagger}_{\mu}\vv{V}^{\mu}\nn\\
  &\ + i g_s \vv{V}_\mu^\dagger T_a \vv{V}_\nu \left( \kappa_s G^{\mu\nu}_a + \widetilde \kappa_s \widetilde G^{\mu\nu}_a \right) \nonumber\\
&\ + i g' Y \vv{V}_\mu^\dagger \vv{V}_\nu \left( \kappa_Y B^{\mu\nu} + \widetilde \kappa_Y  \widetilde B^{\mu\nu}\right)\nn\\
&\ +ig\vv{V}_\mu^\dagger \frac{\tau_{a}}{2} \vv{V}_\nu \left(\kappa_{w}W_{a}^{\mu\nu} + \widetilde{\kappa}_{w}\widetilde{W}_{a}^{\mu\nu}\right).
\end{align}
It interacts with the electromagnetic field $A^{\mu}$ like the singlet vLQ [see Eq. \eqref{eq: singlet photon}], except $\kappa_{Y}$ and $\widetilde{\kappa_{Y}}$ are replaced by effective couplings:
\begin{align}
    \kappa_{eff}=\frac{1}{Q}\left(Y\kappa_Y+T_{3}\kappa_w\right), \quad \widetilde{\kappa}_{eff}=\frac{1}{Q}\left(Y\widetilde{\kappa}_Y+T_{3}\widetilde{\kappa}_w\right).
\end{align}
We set $\kappa_{eff}=1$ and $\widetilde{\kappa}_{eff}=0$ in our calculations.
As shown in Table~\ref{tab:vLQYuk}, two weak-doublet representations are possible: $V_{2}(\mathbf{\overline{3}},\mathbf{2},{5/6})$ and $\widetilde{V}_{2}(\mathbf{\overline{3}},\mathbf{2},{-1/6})$.


The interactions of the $V_2 = \{V_2^{4/3}, V_2^{1/3}\}$ (where the superscripts indicate the electric charges of the components) are shown in Table~\ref{tab:vLQYuk}.
Both $V_2^{1/3}$ and $V_2^{4/3}$ contribute to $a_{\m}$ (see Appendix~\ref{sec:generalgM2}). Since $V_{2}^{4/3}$ has both left-handed and right-handed interactions, it can positively contribute to the chirality-flipping term in $a_\m$ when the product $x^{LR}_2x^{RL}_2$ is positive. The $V_2^{1/3}$ component only gives a chirality-preserving positive contribution to $a_\mu$.

The interactions for the $ \widetilde{V}_{2} = \{\widetilde{V}_2^{1/3}, \widetilde{V}_2^{-2/3}\}$ are shown in Table~\ref{tab:vLQYuk}.
Since $\widetilde{V}_2^{-2/3}$ does not couple to muon, it is only the $\widetilde{V}_2^{1/3}$ that contributes to $a_\m$.  
This contribution is always negative; hence, the $\widetilde V_2$ can not explain the positive shift needed to explain the $(g-2)_\mu$ anomaly.

\subsection{Weak-triplet vector leptoquark}
\noindent 
The interactions and mass terms for a generic weak-triplet vLQ can be expressed as
\begin{align}\label{eq:triplet gauge}
    \mathcal{L} \supset&\  -\frac{1}{2} Tr [\vv{V}_{\mu\nu}^\dagger \vv{V}^{\mu\nu}]+ M_{V}^{2}\vv{V}^{\dagger}_{\mu}\vv{V}^{\mu}\nn\\
    &\ + i g_s \vv{V}_\mu^\dagger T_a \vv{V}_\nu \left( \kappa_s G^{\mu\nu}_a + \widetilde \kappa_s \widetilde G^{\mu\nu}_a \right) \nonumber\\
    &\ + i g' Y \vv{V}_\mu^\dagger \vv{V}_\nu \left( \kappa_Y B^{\mu\nu} + \widetilde \kappa_Y  \widetilde B^{\mu\nu}\right)\nn\\
   &\ +ig\vv{V}_{\mu}^\dagger I^{a} \vv{V}_{\nu} \left(\kappa_{w}W_{a}^{\mu\nu} + \widetilde{\kappa}_{w}\widetilde{W}_{a}^{\mu\nu}\right),
\end{align}
where $I^{a}_{ij}=-i\epsilon_{ij}^a$ are the generators of the $SU(2)$ group in the adjoint representation.  There is only one triplet vLQ species, $U_3(\mathbf{3},\mathbf{3},2/3)$, as shown in 
Table~\ref{tab:vLQYuk}.

Table~\ref{tab:vLQYuk} shows the interactions of the $U_3 = \{U_{3}^{5/3},U_3^{-1/3}$ and $U_{3}^{2/3}\}$. Only $U_{3}^{2/3}$ and $U_{3}^{5/3}$ components contribute to $a_{\m}$. Since for $\kappa_{eff}=1$ and $\widetilde{\kappa}_{eff}=0$, both contribute positively, they can lead to a positive shift in $a_\mu$. However, in the absence of any chirality-flipping contribution, $x^{LL}_3$ has to be large to explain the $(g-2)_\mu$ anomaly. 

\begin{table}[b!]
\caption{vLQs that can produce the necessary $\Delta a_{\ell}$ in the single-coupling scenarios. However, in all the single coupling scenarios, the values of the couplings exceed the perturbative limit (see Table~\ref{tab:dipmom1coup}). For $U_1$ and $V_2$, the \cmark~ in parenthesis indicates that the necessary $\Delta a_{\ell}$ value can be achieved in two-coupling scenarios with perturbative couplings through chirality-flipping terms (see Table~\ref{tab:dipmom1coup} and Figs.~\ref{fig:gM2} and~\ref{fig:electronAMDM}).
\label{tab:vLQgm2}}
{\renewcommand\baselinestretch{1.25}\selectfont
\begin{tabular*}{\columnwidth}{l @{\extracolsep{\fill}}ccccc}
\hline\hline
 & $ U_{1}$ & $\widetilde{U}_{1}$  & $V_{2}$ & $\widetilde{V}_{2}$ &  $U_{3}$\\
\hline
$\Delta a_\m$ & \cmark~(\cmark) & \cmark & \xmark~(\cmark) & \xmark & \cmark\\
$\Delta a_e^{Rb}$ & \cmark~(\cmark) & \cmark & \xmark~(\cmark) & \xmark & \cmark\\
$\Delta a_e^{Cs}$ & \xmark~(\cmark) & \xmark & \cmark~(\cmark) & \cmark & \xmark\\
\hline\hline
\end{tabular*}}
\end{table}

\section{Fitting the $(g-2)_\ell$ measurements}\label{sec:gm2fit}
\noindent
We perform parameter scans to obtain the parameter regions consistent with the experimental values of $(g-2)_\ell$. 
The significant contributions come from the chirality-flipping terms that depend on the product of two independent couplings (which, in general, could be complex) as ${\rm Re} (x^{L\al}_{i\ell}{x^{R\overline\alpha}_{i\ell}}^*)$ [see Eq.~\eqref{eq:LQAMDM}]. Since the LHC only limits the absolute values of these couplings, we assume all couplings to be real for simplicity and vary the couplings within the perturbative range, i.e., $\lt[-\sqrt{4\pi},\sqrt{4\pi}\rt]$. In most plots, however, we do not show the entire range but focus only on the interesting parts. 

We list the vLQs that can, in principle, provide a positive shift to $(g-2)_\mu$, i.e., $\Dl a_\mu > 0$ in the limit $\kappa_{Y}=1$, $\widetilde{\kappa}_{Y}=0$ (or $\kappa_{eff}=1$, $\widetilde{\kappa}_{eff}=0$) in Table~\ref{tab:vLQgm2}. However, if we take a bottom-up minimalist point of view in terms of the number of new vLQ couplings, we find that no vLQ can explain the $(g-2)_\mu$ anomaly with just one coupling of $\mc{O}(1)$ or smaller (while the rest are practically zero or negligible). Table~\ref{tab:dipmom1coup} shows the ranges of the individual couplings needed to accommodate the $(g-2)_\ell$ measurements at $1\sg$ level for $M_{\ell_q}=2.5$~TeV. With just one coupling, only the chirality-preserving term contributes. Since it is suppressed by $m_\mu^2/M_{\ell_q}^2$, the associated coupling must be pretty large to explain the anomaly. The coupling can be lower for lighter vLQs; however, since the current bounds on vLQ masses from the direct pair-production searches are roughly about $2$~TeV, the vLQs cannot be much lighter. On the other hand, increasing the mass would only push the coupling to a bigger value beyond the perturbative regime. 

Only $U_1$ and $V_2$ can accommodate the $(g-2)_\mu$ anomaly with two non-negligible couplings through the chirality-flipping term $(\sim m_q m_\mu)$ (see Table~\ref{tab:dipmom2coup}). There are three possible two-coupling scenarios for the three generations of quarks.  For the third-generation quarks, the chirality-flipping term dominates, but for the lighter quarks, the chirality-preserving term dominates. Hence, in the $\{x^{LL}_{32}, x^{RR}_{32}\}$ scenario for $U_1$ and the $\{x^{LR}_{32}, x^{RL}_{32}\}$ scenario for $V_2$, the couplings are much smaller than the light-quark couplings to address the $(g-2)_\mu$ anomaly. (Since now we focus only on $U_1$ and $V_2$, we suppress the subscript indicating the vLQ representations from the couplings, as it is clear from the context; see Table~\ref{tab:vLQYuk}). 
Among the up- and down-aligned scenarios, the up-aligned $x_{32}$ scenarios for $U_1$ and $V_2$ can contribute to the $R_K^{(*)}$ observable through quark mixing. Since the $R_K^{(*)}$ in the SM is a loop-induced process, but vLQs contribute at the tree level, the $R_K^{(*)}$ measurements eliminate almost the entire parameter space. This makes the up-aligned $x_{32}$ scenarios for $U_1$ and $V_2$ very restrictive. Hence, we do not consider them further in our analysis. 

If we consider the electron couplings instead of the muon, we get the vLQ contributions to electron AMDM. As mentioned earlier, $\Dl a_e^{Rb}$ is positive whereas $\Dl a_e^{Cs}$ is negative. Hence, essentially $U_1$, $\widetilde{U}_1$, and $U_3$ can accommodate the $\Dl a_e^{Rb}$ measurement; whereas parameter regions consistent with the $\Dl a_e^{Cs}$ measurement can only be found for $V_2$ and $\widetilde{V}_2$ (see Table~\ref{tab:vLQgm2})---there are no common overlapping regions. We show the preferred coupling ranges for scenarios with one electron coupling and two electron couplings obtained using $\Dl a_e^{Rb}$ and $\Dl a_e^{Cs}$ measurements. For two couplings, the allowed coupling values are small when vLQs are connected with the first- and second-generation quarks. Therefore, these scenarios may agree with the LHC data.

As discussed in Sec.~\ref{sec:dipmomlep}, the current sensitivity on the measurements of $a_\tau$ is poor. Therefore, the experimental measurements of $a_\tau$ cannot set any significant limit on the vLQ parameter space. In the future, if the sensitivity on $a_\tau$ improves to $\sim 10^{-6}$, one might constrain vLQ couplings within the perturbative range especially involving third-generation quarks. A similar analysis for scalar LQ has been performed in Ref.~\cite{Crivellin:2021spu}.

\begin{table*}
\caption{\label{tab:dipmom1coup} 
Summary of the bounds on the vLQ couplings (listed in Table~\ref{tab:vLQYuk}) contributing to the AMDMs of the electron and muon through the chirality-preserving term [$\sim m_\ell^2/M_{\ell_q}^2$, see Eq.~\eqref{eq:LQAMDM}], considering each coupling separately (single-coupling scenarios). These are $1\sg$ bounds on the absolute values of the couplings for $M_{\ell_q} = 2.5$~TeV.}
\centering{\footnotesize\renewcommand\baselinestretch{2}\selectfont
\begin{tabular*}{\textwidth}{l @{\extracolsep{\fill}}ccccc}
\hline \hline 
vLQ & Couplings & $Q_q$ & From $\Delta a_\mu$ &  From $\Delta a_e^{Cs}$ &  Form $\Delta a_e^{Ru}$ \\
\hline
$U_1^{2/3}$& $|x_{1\ell}^{LL}|,|x_{1\ell}^{RR}|$ &$-\frac{1}{3} $& $[6.18,7.62]$ & ---&$ [12.56,25.77]$\\
$\widetilde{U}_1^{5/3}$& $|x_{1\ell}^{RR}|$&$\frac{2}{3}$ & $[5.65,7.10]$&--- & $[11.25,23.40]$\\
$V_2^{1/3},V_2^{4/3}$& $|x_{2\ell}^{LR}|,|x_{2i}^{RL}|$&$\frac{2}{3},-\frac{1}{3}$ & ---& $[17.20,26.73]$ &---\\
$\widetilde{V}_2^{1/3}$& $|x_{2\ell}^{RL}|$& $\frac{2}{3}$&---& $[26.04,40.75]$ & ---\\
$U_3^{2/3}$&$|x_{3\ell}^{LL}|$ &$-\frac{1}{3}$ &$[3.30,4.11]$ &--- &$[6.65,14.15]$ \\
\hline\hline
\end{tabular*}}
\end{table*}

\begin{table*}
\caption{\label{tab:dipmom2coup} Similar to Table~\ref{tab:dipmom1coup} but for product of couplings in the $U_1$ and $V_2$ up- and down-aligned two-coupling scenarios. In these scenarios, the chirality-flipping term [$\sim m_\ell m_q/M_{\ell_q}^2$, see Eq.~\eqref{eq:LQAMDM}] contributes dominantly.}
\centering{\footnotesize\renewcommand\baselinestretch{2}\selectfont
\begin{tabular*}{\textwidth}{@{\extracolsep{\fill}}cccccc}
\hline \hline 
        vLQ          &          Coupling products         & Alignment &  From $\Delta a_{\mu}$ &  From $\Delta a_e^{Cs}$ &  From $\Delta a_e^{Ru}$ \\ \hline
\multirow{6}{*}{$U_1$} & \multirow{2}{*}{$x_{3\ell}^{LL}x_{3\ell}^{RR}$} & up &$ [-0.55,-0.37]$ & $[0.02,0.04]$ & $[-0.03,-0.007]$ \\  
                  &                   & down & $[-0.55,-0.37]$ & $[0.02,0.05]$ & $[-0.03,-0.007]$ \\ 
                  & \multirow{2}{*}{$x_{2\ell}^{LL}x_{2\ell}^{RR}$} & up & $[-8.70,-5.88]$ &$[0.31,0.77] $ &$[-0.50,-0.11]$  \\  
                  &                   & down & $[-24.35,-16.45]$ & $[0.88,2.14]$ & $[-1.40,-0.31]$ \\ 
                  & \multirow{2}{*}{$x_{1\ell}^{LL}x_{1\ell}^{RR}$} & up & $[-56.67,-38.30]$ &$[2.13,5.07]  $& $[-3.20,-0.73]$ \\  
                  &                   & down & $[-488.0,-329.0]$ &$[18.0,43.4] $ & $[-27.7,-6.4]$ \\ \hline
\multirow{6}{*}{$V_2$} & \multirow{2}{*}{$x_{3\ell}^{LR}x_{3\ell}^{RL}$} & up & $[0.22,0.33]$  & $[-0.03,-0.01]$ & $[0.004,0.018]$ \\ 
                  &                   & down & $[0.22,0.33]$ & $[-0.03,-0.01]$ & $[0.004,0.018]$ \\ 
                  & \multirow{2}{*}{$x_{2\ell}^{LR}x_{2\ell}^{RL}$} & up & $[3.54,5.31]$ &$[-0.46,-0.19]$  & $[0.06,0.30]$ \\ 
                  &              & down & $[9.87,14.80]$ &$[-1.30,-0.54]$  &$[0.18,0.80]  $\\ 
                  & \multirow{2}{*}{$x_{1\ell}^{LR}x_{1\ell}^{RL}$} & up & $[23.40,34.12]$ &$[-2.96,-1.26]$  & $[0.43,1.86]$ \\ 
                  &              & down & $[201.52,296.65]$ & $[-26.0,-11.0]$ & $[3.80,16.11]$ \\ \hline\hline
\end{tabular*}}
\end{table*}

\begin{figure}[!t]
\centering
\captionsetup[subfigure]{labelformat=empty}
\subfloat[\quad\quad\quad(a) Up/down aligned]{
\includegraphics[width=0.32\textwidth,height=0.32\textwidth]{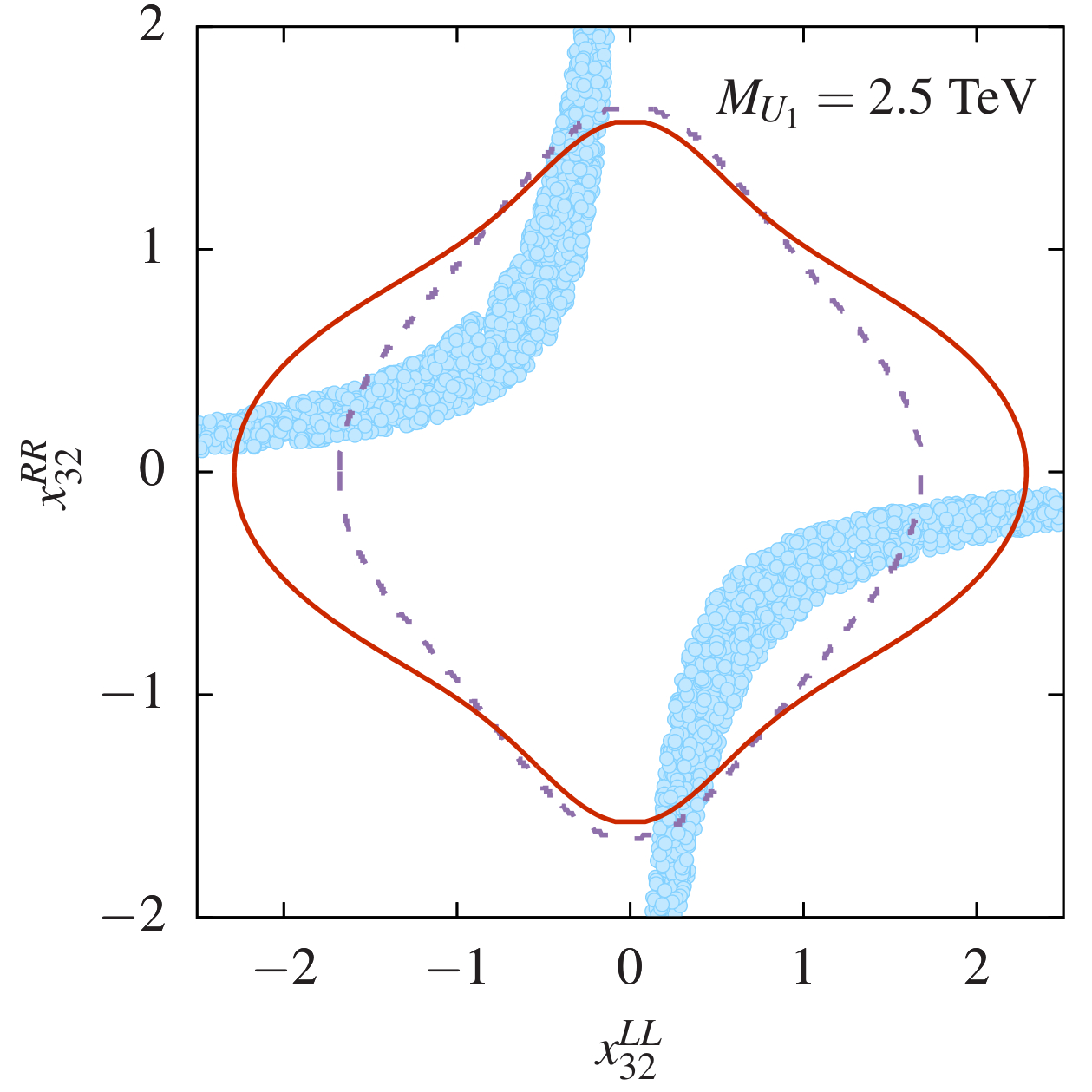}\label{fig:gM2a}}\\
\subfloat[\quad\quad\quad(b) Up/down aligned]{
\includegraphics[width=0.32\textwidth,height=0.32\textwidth]{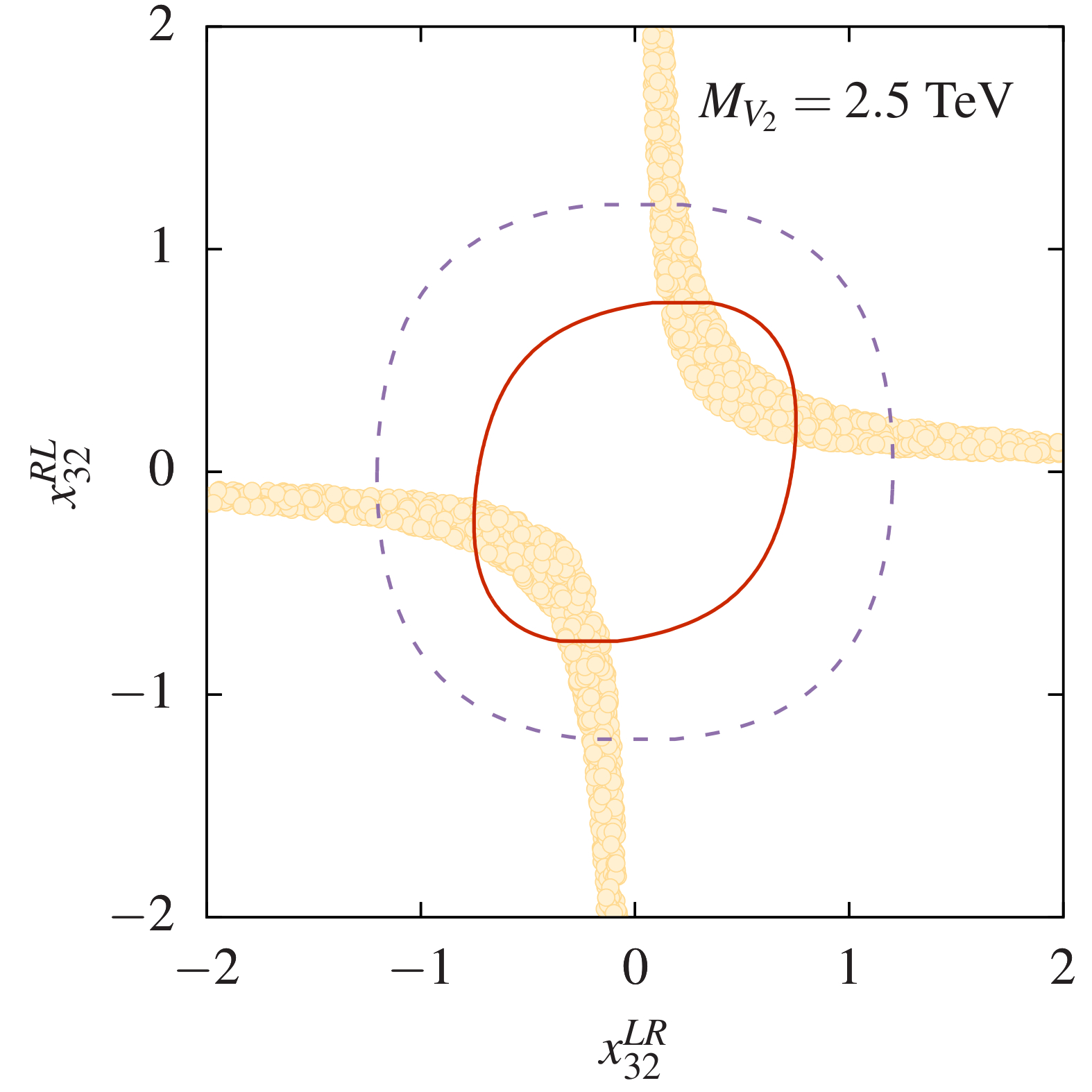}\label{fig:gM2b}}
\caption{\label{fig:gM2}$U_1$ and $V_2$ $2\sigma$ $(g-2)_\m$ parameter spaces and the corresponding LHC direct search (solid) and indirect dilepton (dashed) bounds. The regions inside the bounds are allowed. The difference between the up- and down-aligned scenarios can be seen from Table~\ref{tab:vLQYuk}.}
\end{figure}
\begin{figure*}
\centering
\captionsetup[subfigure]{labelformat=empty}
\subfloat[\quad\quad\quad(a) Down aligned, $\prescript{133}{55}{\mathbf{Cs}}$ exp.]{\includegraphics[width=0.275\textwidth,height=0.275\textwidth]{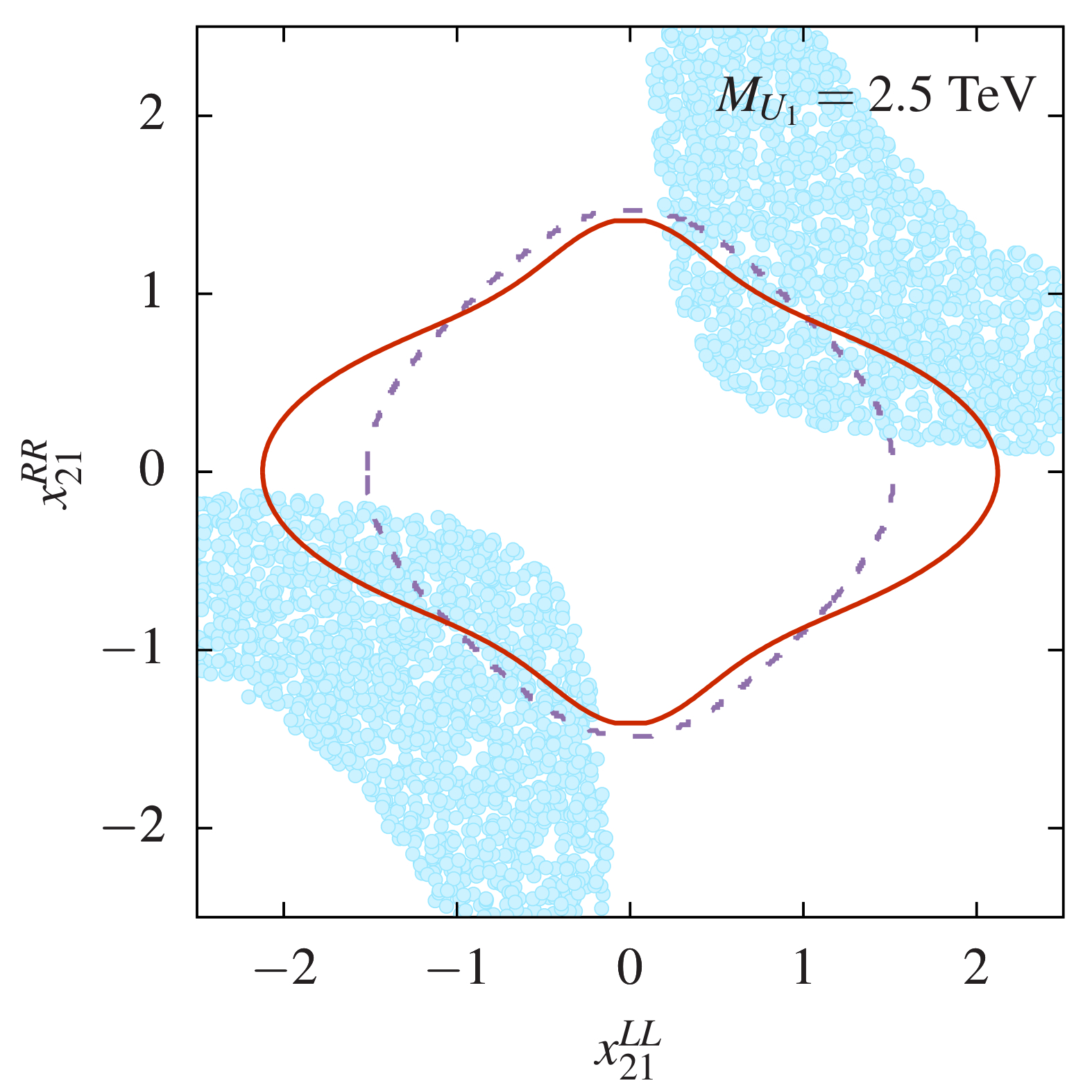}\label{fig:3a}}\hspace{1cm}
\subfloat[\quad\quad\quad(b) Up aligned, $\prescript{133}{55}{\mathbf{Cs}}$ exp.]{\includegraphics[width=0.275\textwidth,height=0.275\textwidth]{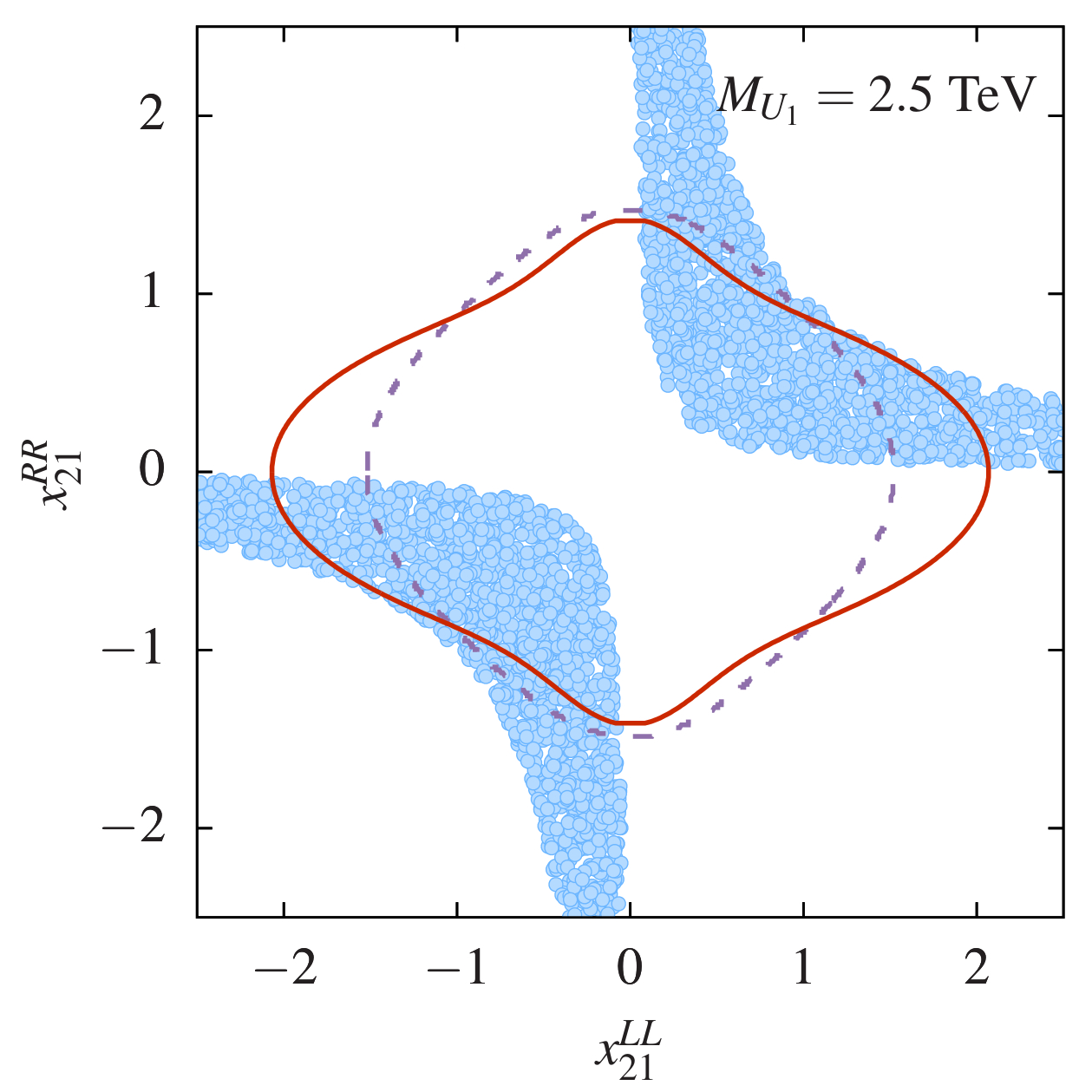}\label{fig:3b}}\hspace{1cm}
\subfloat[\quad\quad\quad(c) Up/down aligned, $\prescript{133}{55}{\mathbf{Cs}}$ exp.]{\includegraphics[width=0.275\textwidth,height=0.275\textwidth]{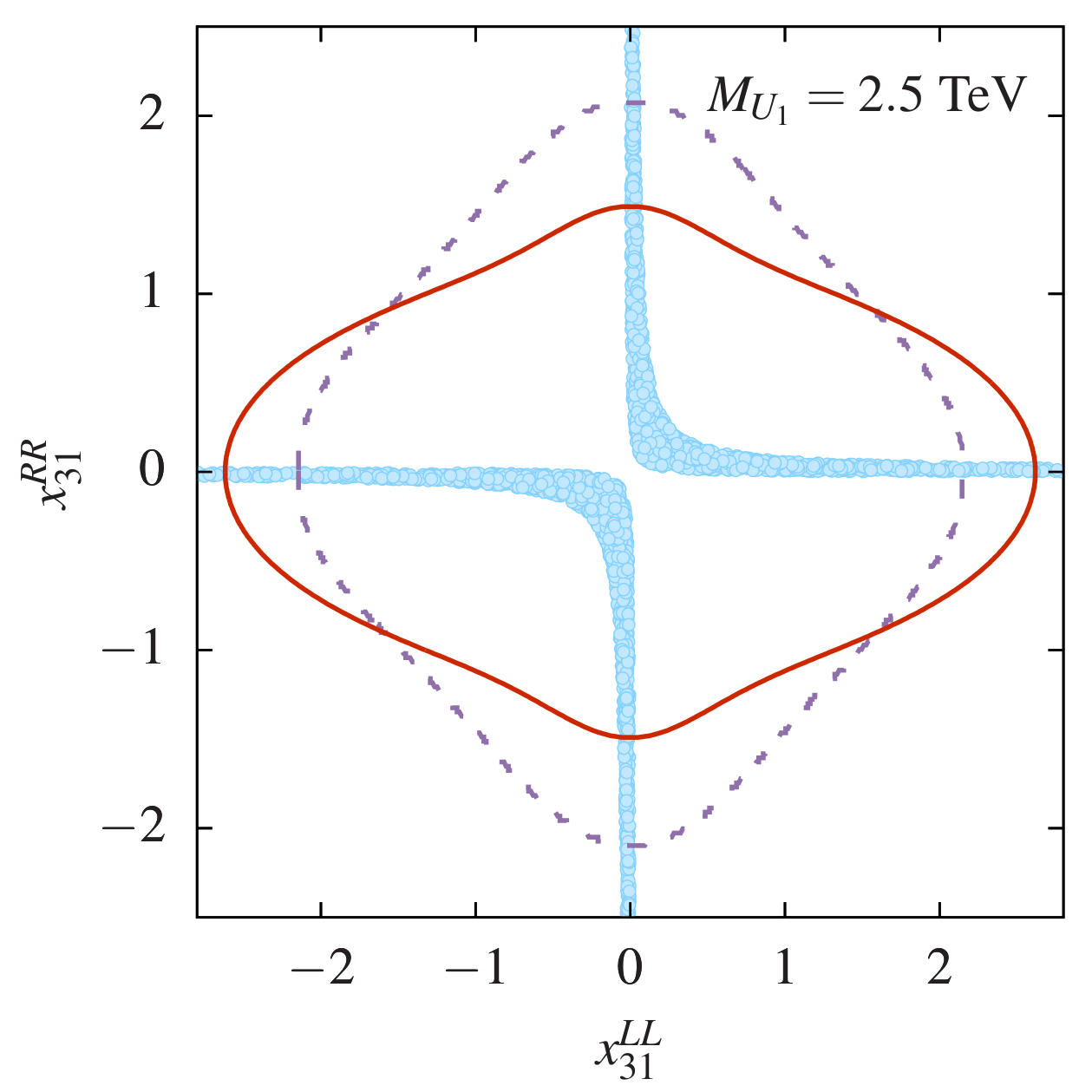}\label{fig:3c}}\\
\subfloat[\quad\quad\quad(d) Down aligned, $\prescript{87}{37}{\mathbf{Ru}}$ exp.]{\includegraphics[width=0.275\textwidth,height=0.275\textwidth]{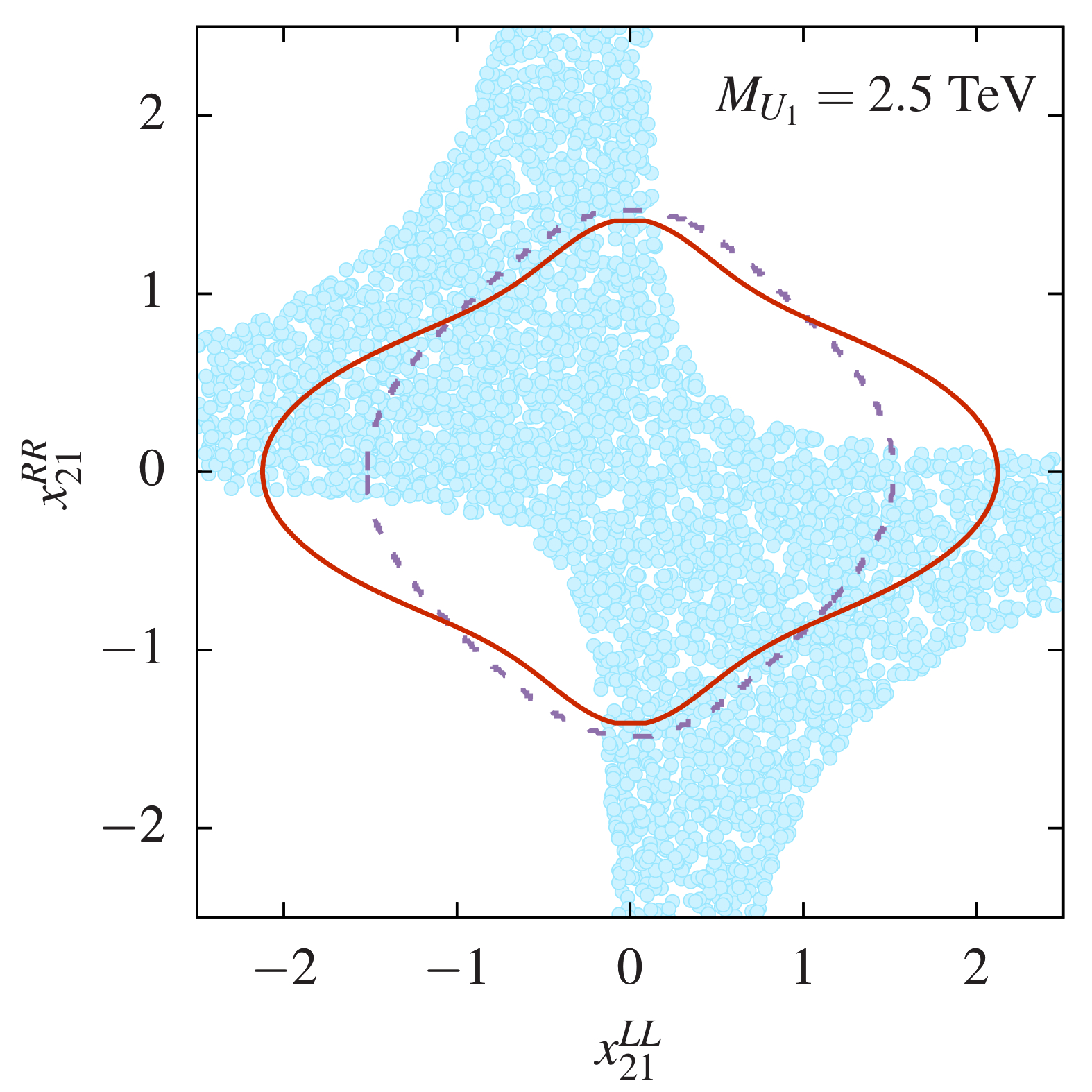}\label{fig:3d}}\hspace{1cm}
\subfloat[\quad\quad\quad(e) Up aligned, $\prescript{87}{37}{\mathbf{Ru}}$ exp.]{\includegraphics[width=0.275\textwidth,height=0.275\textwidth]{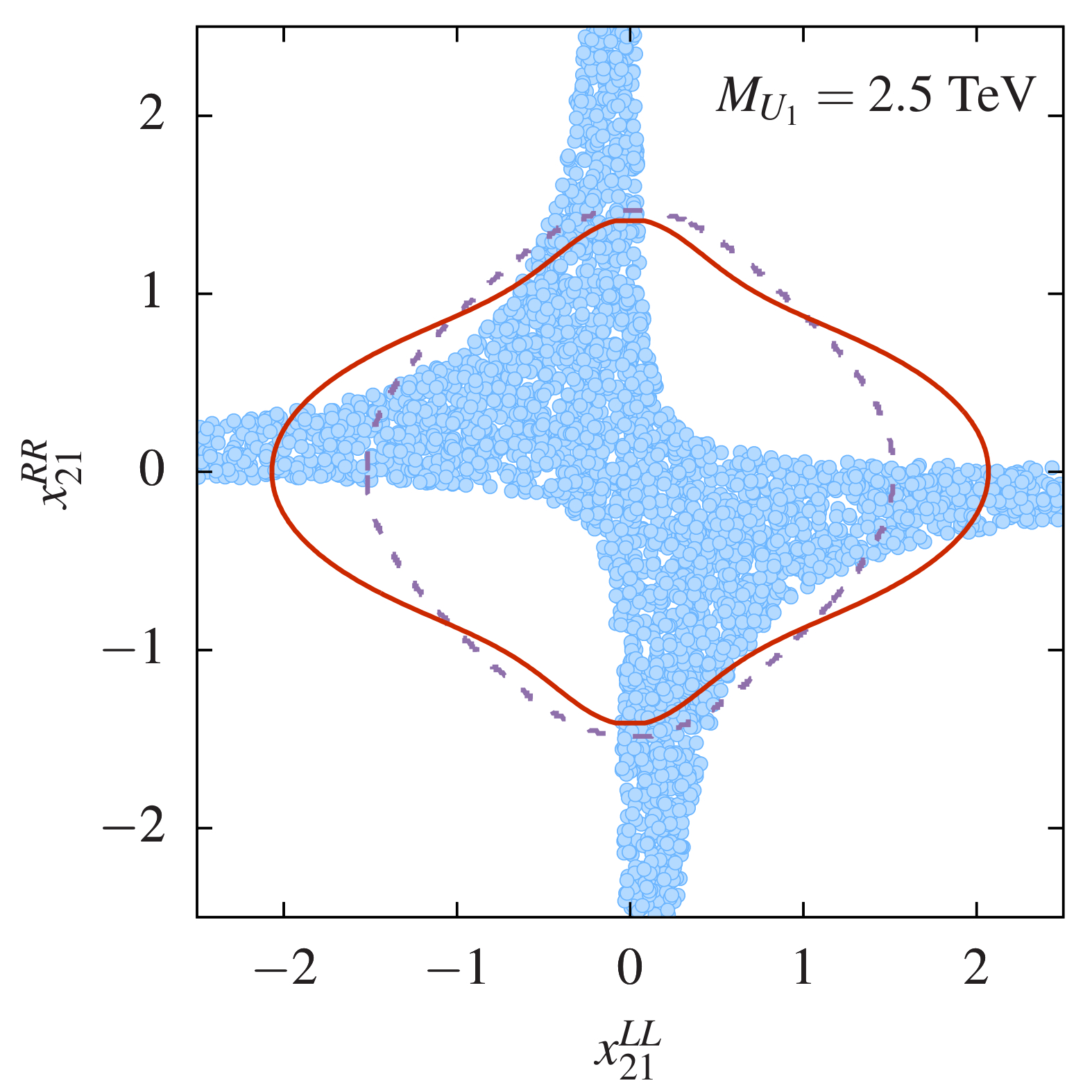}\label{fig:3e}}\hspace{1cm}
\subfloat[\quad\quad\quad(f) Up/down aligned, $\prescript{87}{37}{\mathbf{Ru}}$ exp.]{\includegraphics[width=0.275\textwidth,height=0.275\textwidth]{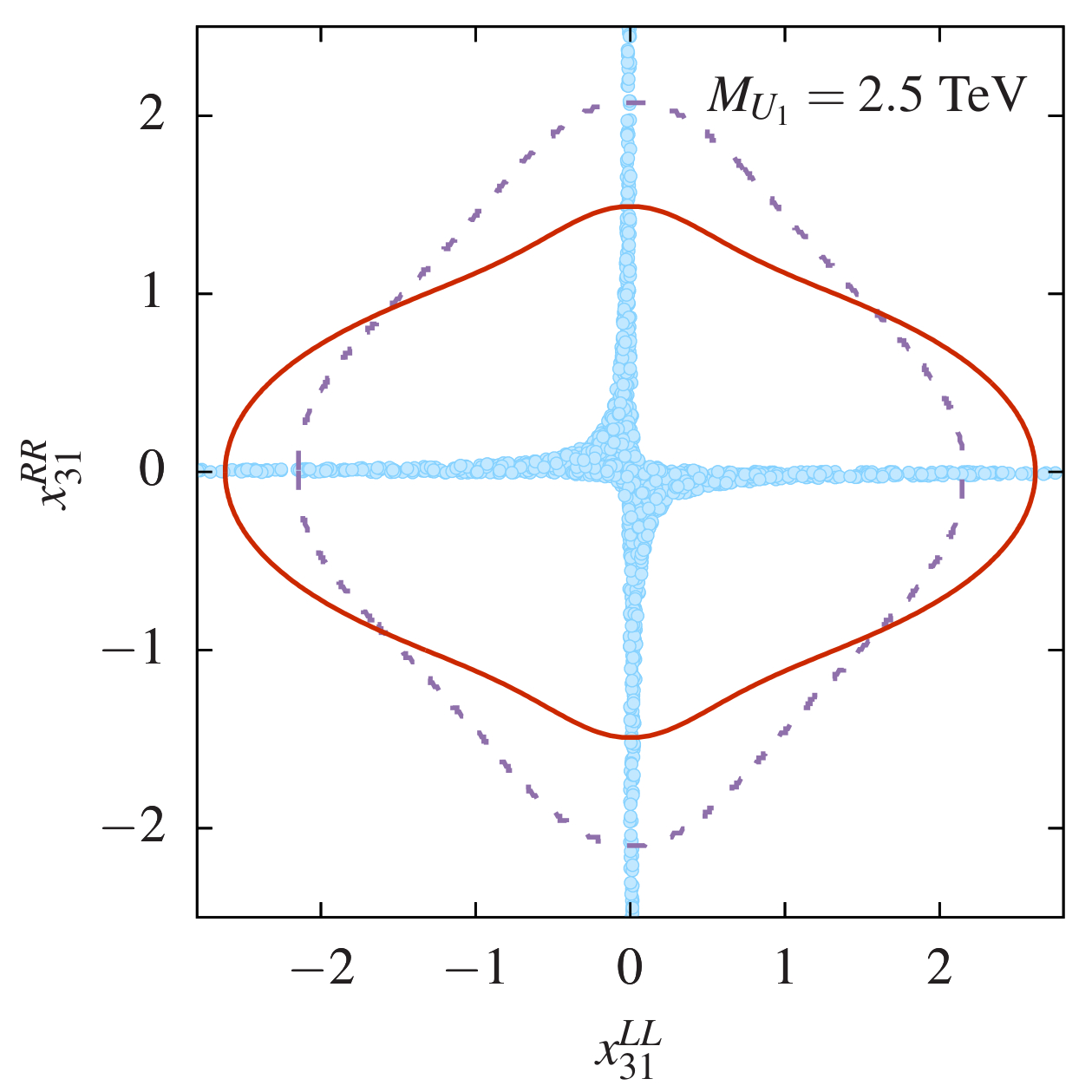}\label{fig:3f}}\\
\subfloat[\quad\quad\quad(g) Down aligned, $\prescript{133}{55}{\mathbf{Cs}}$ exp.]{\includegraphics[width=0.275\textwidth,height=0.275\textwidth]{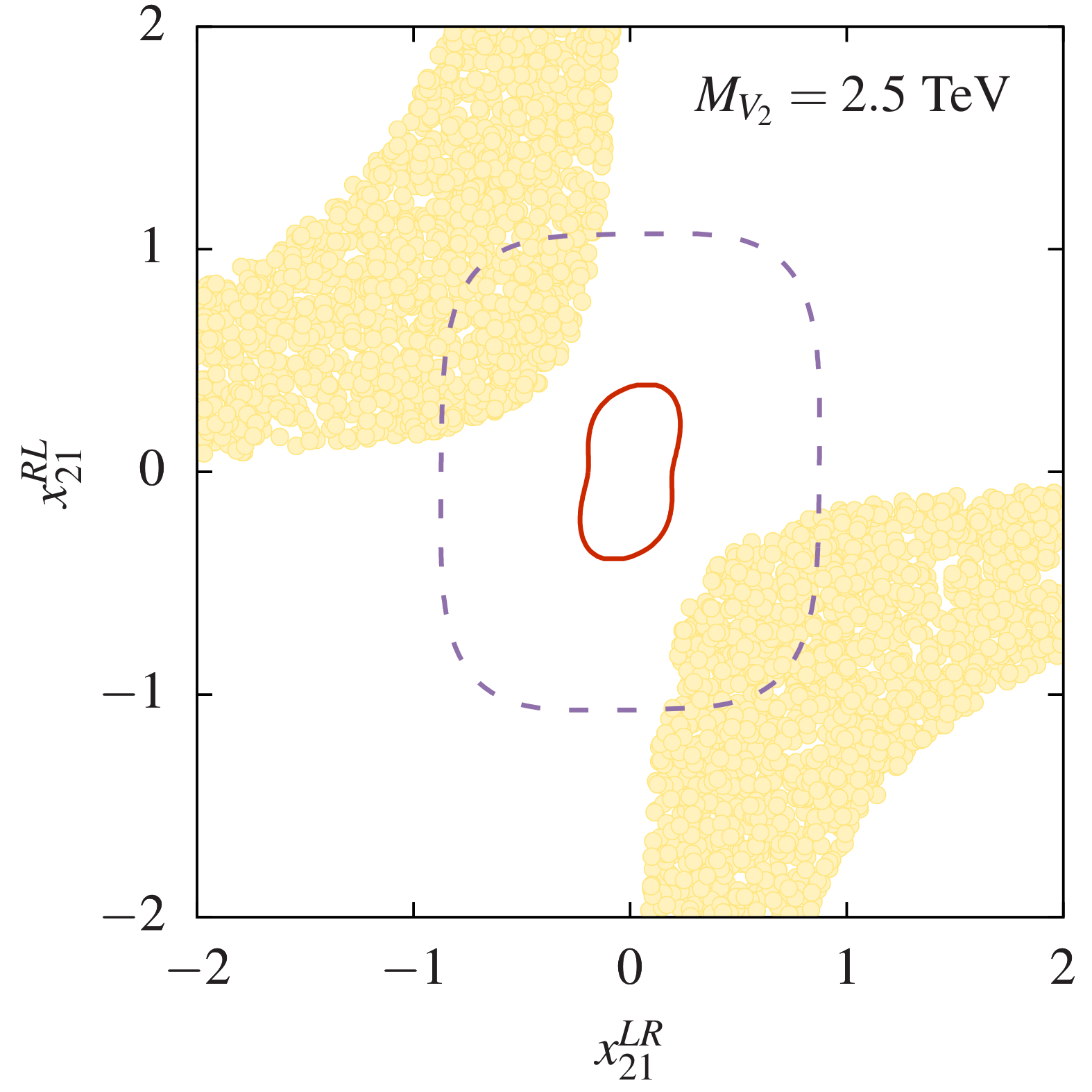}\label{fig:3g}}\hspace{1cm}
\subfloat[\quad\quad\quad(h) Up aligned, $\prescript{133}{55}{\mathbf{Cs}}$ exp.]{\includegraphics[width=0.275\textwidth,height=0.275\textwidth]{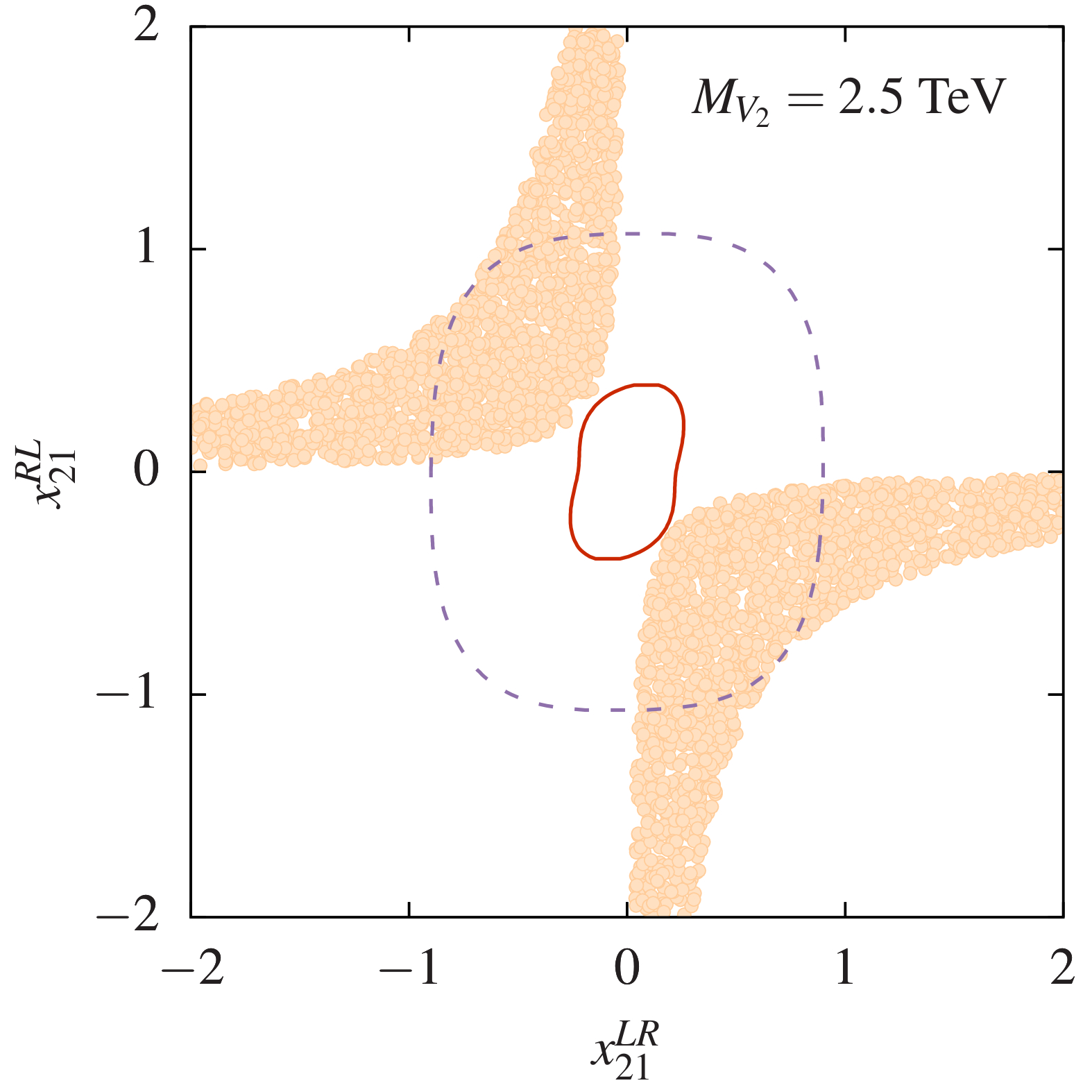}\label{fig:3h}}\hspace{1cm}
\subfloat[\quad\quad\quad(i) Up/down aligned, $\prescript{133}{55}{\mathbf{Cs}}$ exp.]{\includegraphics[width=0.275\textwidth,height=0.275\textwidth]{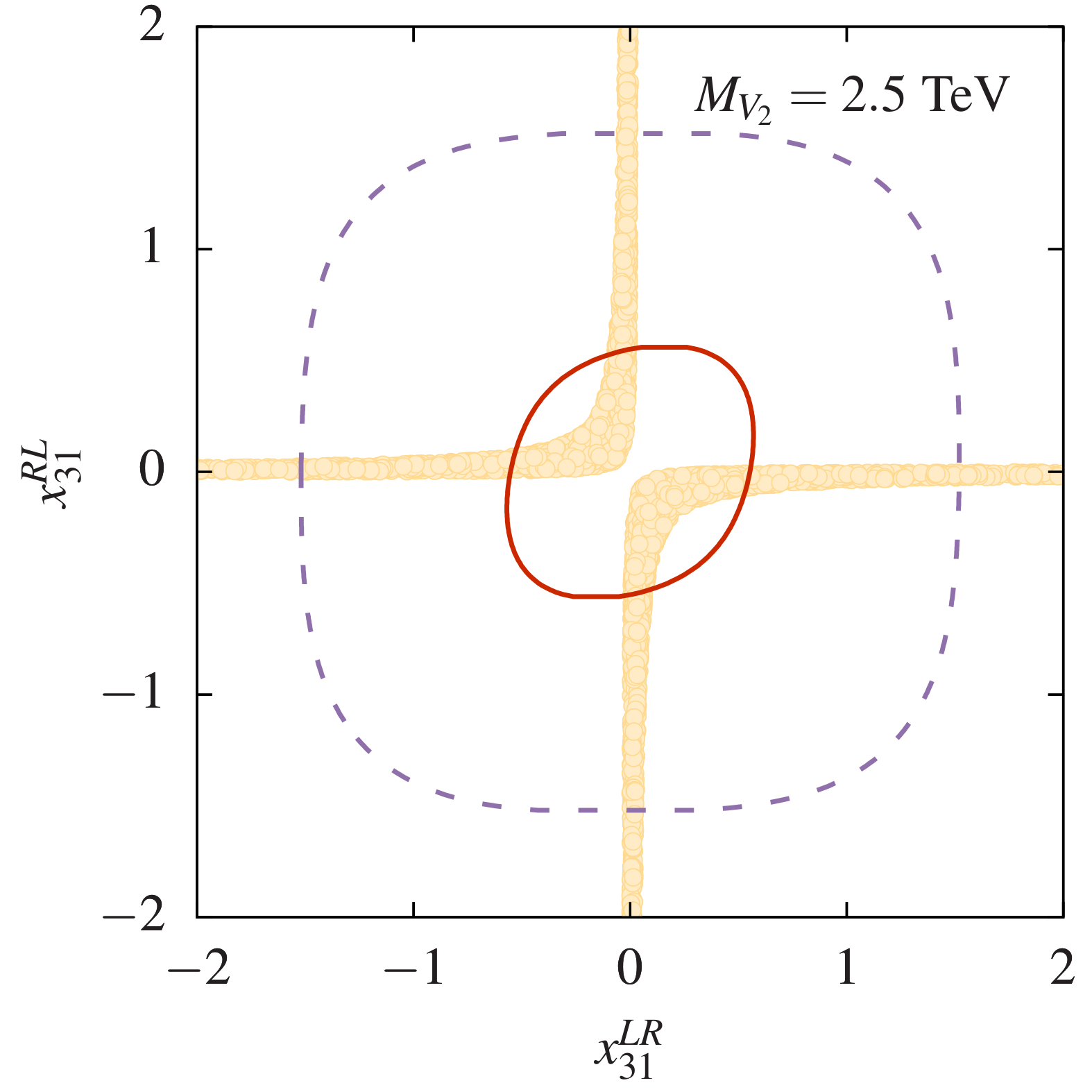}\label{fig:3i}}\\
\subfloat[\quad\quad\quad(j) Down aligned, $\prescript{87}{37}{\mathbf{Ru}}$ exp.]{\includegraphics[width=0.275\textwidth,height=0.275\textwidth]{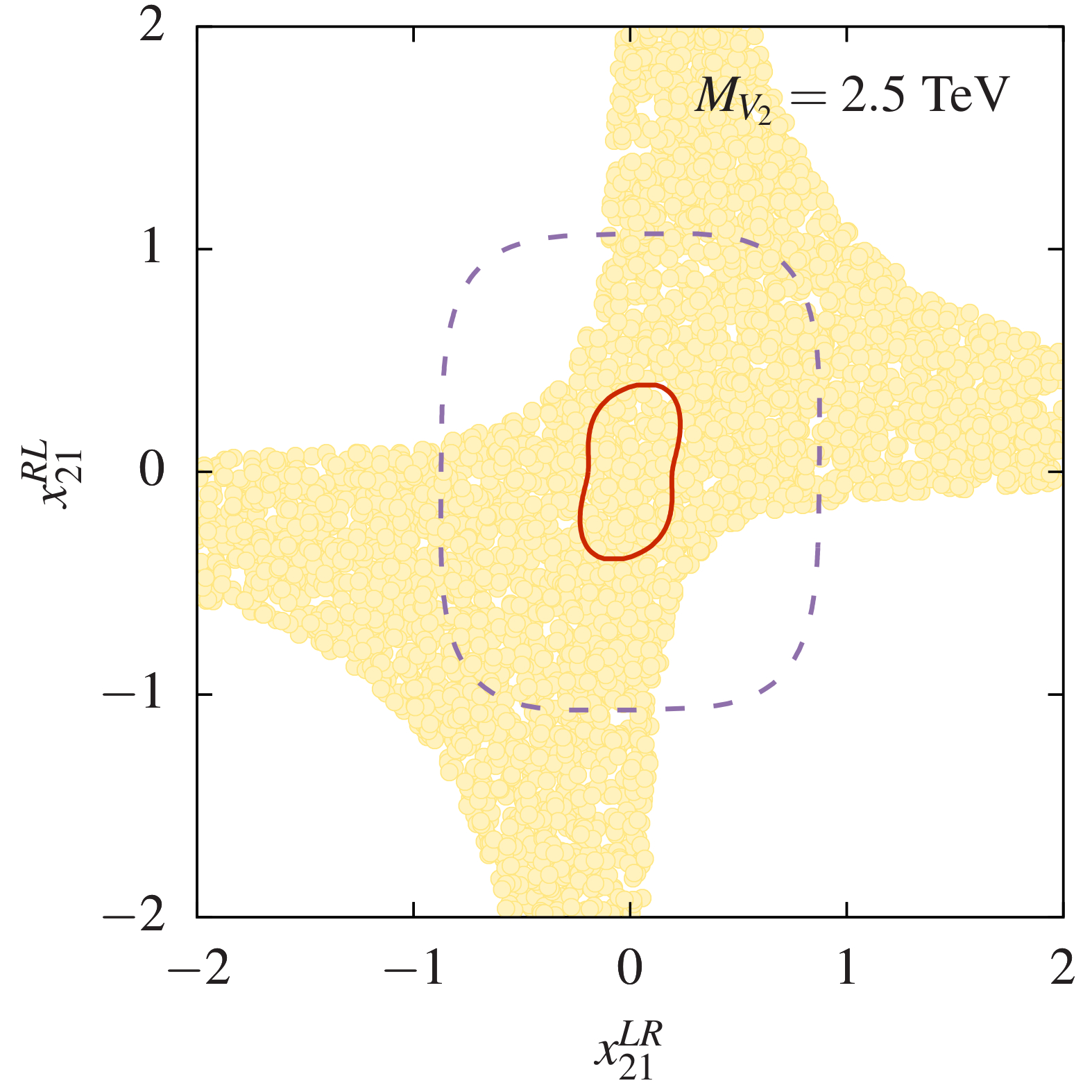}\label{fig:3j}}\hspace{1cm}
\subfloat[\quad\quad\quad(k) Up aligned, $\prescript{87}{37}{\mathbf{Ru}}$ exp.]{\includegraphics[width=0.275\textwidth,height=0.275\textwidth]{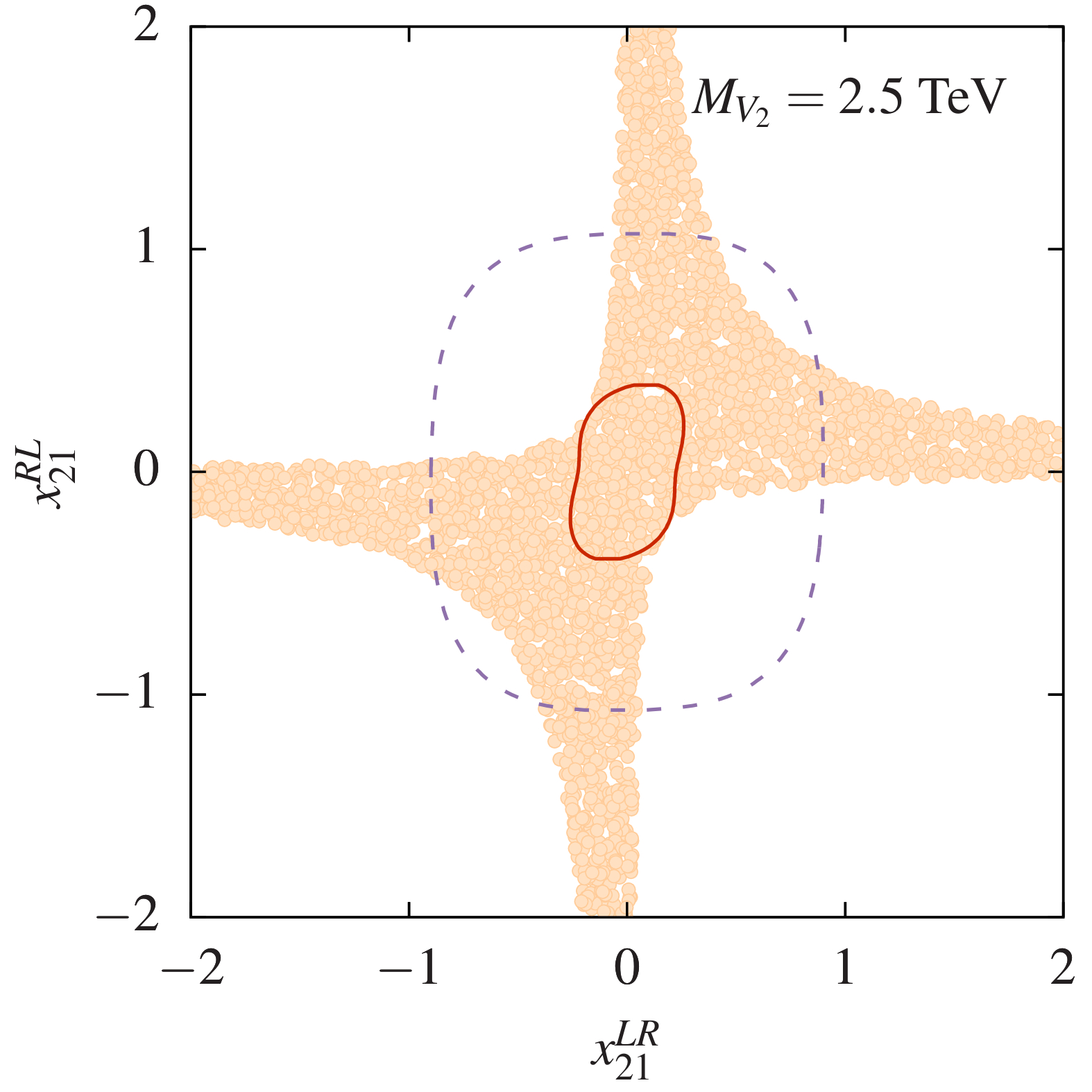}\label{fig:3k}}\hspace{1cm}
\subfloat[\quad\quad\quad(l) Up/down aligned, $\prescript{87}{37}{\mathbf{Ru}}$ exp.]{\includegraphics[width=0.275\textwidth,height=0.275\textwidth]{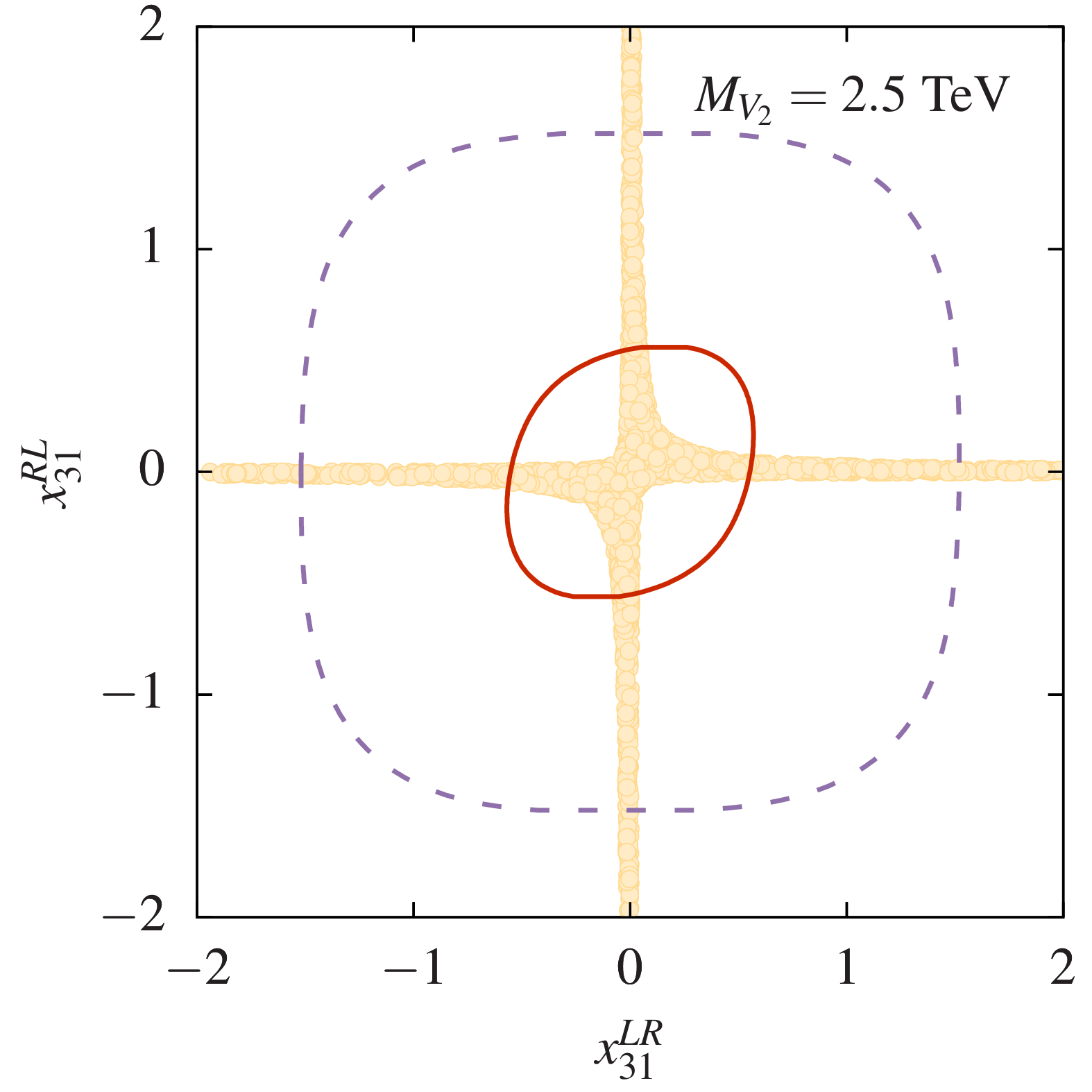}\label{fig:3l}}\\
\caption{Same as Fig.~\ref{fig:gM2} except for the electron instead of the muon.}
\label{fig:electronAMDM}
\end{figure*}

\section{LHC constraints and results}
\label{sec:results}
\noindent
It is well-known that the high-energy LHC data strongly constrain the parameter spaces of vLQs, especially in large-coupling limits. Because of the relatively larger parton density functions (PDFs) of the light quarks, the LHC constraints on vLQ couplings are particularly stringent when the vLQs interact with the first- and second-generation quarks~\cite{Bhaskar:2021pml}. On the other hand, as Tables~\ref{tab:dipmom1coup} and \ref{tab:dipmom2coup} indicate, large couplings are required to satisfy the observed shift in $a_\mu$. Consequently, there is no overlap between the $(g-2)_\mu$ parameter regions and those allowed by the LHC data. However, when vLQs couple with third-generation quarks, overlapping regions open up.

We briefly discuss how one can recast the LHC data to constrain the LQ parameter space~\cite{Bhaskar:2023ftn}. The LHC limits are of two types---direct (mainly from the LQ pair production searches in the dilepton-dijet channels~\cite{CMS:2018qqq,*CMS:2020wzx,*ATLAS:2022wcu}) and indirect (from the dilepton resonance searches~\cite{CMS:2021ctt}). 
\begin{enumerate}
\item[(a)] 
\emph{Dilepton-dijet ($\ell\ell jj$) data:} the current limit on vLQ mass from the pair-production searches is around $2$ TeV for $100\%$ decay in the search mode. The limits weaken when the branching ratio (BR) in the search mode goes down in scenarios with multiple couplings. However, as illustrated in Ref.~\cite{Bhaskar:2023ftn} (earlier in Refs.~\cite{Mandal:2015vfa,Mandal:2012rx,*Mandal:2016csb,*Bhaskar:2023xkm} in related contexts), other production processes---such as single productions, indirect productions ($t$-channel vLQ exchanges and their interference with SM processes)---can also contribute to the $\ell\ell jj$ final states. These additional contributions become significant for order-one vLQ couplings, leading to stronger mass exclusion bounds. We obtain the recast direct limits relevant to our study following the method explained in Ref.~\cite{Bhaskar:2023ftn}.

\item[(b)]\emph{Dilepton ($\ell\ell$) resonance data:}
all the above-mentioned production processes also contribute to the $\ell\ell$ resonance-search signals and affect the high-$p_{\rm T}$ tails of the dilepton distributions. The $t$-channel vLQ exchange process can significantly interfere with the SM background ($q\overline q\to Z/\gm^* \to \ell^+\ell^-$). For example, in the presence of muon couplings, vLQs produced at the LHC would contribute to the $\m\m(+X)$ final state and affect the high-$p_T$ tail of the dimuon distribution observed at the LHC. One can fit the observed dimuon invariant mass distribution from Ref.~\cite{CMS:2021ctt} with vLQ events and estimate the limits on the coupling(s) using the $\chi^2$ estimation technique. The method explained in Ref.~\cite{Bhaskar:2021pml} (also see~\cite{Mandal:2018kau, Aydemir:2019ynb}) is easily extendable to scenarios with multiple new couplings. In this paper, we use an automated implementation of this method~\cite{CaLQ} to locate the region(s) of parameter space allowed by the LHC data in each scenario we consider. In the automated implementation, the vLQ interactions are first implemented in \textsc{FeynRules}~\cite{Alloul:2013bka} to create the Universal FeynRules Output~\cite{Degrande:2011ua} model files for \textsc{MadGraph}~\cite{Alwall:2014hca}. \textsc{Pythia8}~\cite{Bierlich:2022pfr} performs parton showering and hadronization on the events generated in \textsc{MadGraph}. The detector simulation is performed with \textsc{Delphes}~\cite{deFavereau:2013fsa}. (We use the same computational setup to estimate the direct limits.)
\end{enumerate}

For $2.5$ TeV $U_1$ and $V_2$, we show how the direct and indirect LHC bounds constrain large values of the couplings on two-couplings planes. In Fig.~\ref{fig:gM2a}, we show the parameter ranges in the $\{x^{LL}_{32},x^{RR}_{32}\}$ up-/down-aligned $U_1$ scenarios that can address the $(g-2)_\m$ anomaly. These two-coupling scenarios of $U_1$ can address the $a_\m$ discrepancy with perturbative couplings, as shown in Table~\ref{tab:dipmom2coup}. The indirect limit is stronger than the direct one along $x^{LL}_{32}$, as the latter is more sensitive to the reduction in BR($U_1\to \mu b$)  ($x^{LL}_{32}$ lets the $U_1$ decay via $U_1\to t\n$ mode). Fig.~\ref{fig:gM2a} shows similar LHC bounds/$(g-2)_\m$-favoured parameter ranges on the $\{x^{LR}_{32},x^{RL}_{32}\}$ plane for $V_2$. In this case, unlike for the $U_1$, the couplings are positively correlated, and the direct bounds are tighter and severely restrict the viable parameter ranges. We show the corresponding plots for electron AMDM measurements with $\prescript{133}{55}{\mathbf{Cs}}$ and $\prescript{87}{37}{\mathbf{Ru}}$ in Fig.~\ref{fig:electronAMDM}. There, we also show the LHC bounds on the scenarios with second-generation quark couplings, i.e., $\{x^{XX}_{21}, x^{XX}_{31}\}$. We see that good parts of the $U_1$ parameter space survive the LHC bounds but for $V_2$, the LHC data completely rule out the $\{x^{LR}_{21},x^{RL}_{31}\}$ up-/down-aligned scenarios preferred by the $\prescript{133}{55}{\mathbf{Cs}}$ measurement.
\begin{table}[b!]
\caption{Limits on the imaginary couplings of a $2.5$ TeV $U_1$ from the current measurements of the EDM of the electron and the neutron (for $\mu$ and $\tau$). We assume one of the couplings ($x_{3\ell}^{LL}$) to be real. The real part of the other coupling ($x_{3\ell}^{RR}$) is chosen to satisfy the AMDM measurements and the LHC bounds. The limits are obtained for the best-fit value of $\beta^{\tilde{G}}_{n}\approx 2\times 10^{-20} {\rm e~cm}$ [see Eq.~\eqref{eq:neutronEDM}].
\label{tab:EDM}} 
{\renewcommand\baselinestretch{1.4}\selectfont
\begin{tabular*}{\columnwidth}{l @{\extracolsep{\fill}}ccr}
\hline\hline
\multirow{2}{*}{Lepton}&\multicolumn{2}{c}{Benchmark choice}&\multicolumn{1}{c}{Upper limit}\\
\cline{2-4} 
& $x_{3\ell}^{LL}$ & ${\rm Re}(x_{3\ell}^{RR})$ & \multicolumn{1}{c}{${\rm Im}(x_{3\ell}^{RR})$} \\\hline
$e~(\prescript{133}{55}{\mathbf{Cs}})$ &$0.5+0i$ & $0.06$  & $5\times 10^{-8}~(d_e)$ \\
$e~(\prescript{87}{37}{\mathbf{Ru}})$ & $0.5+0i$ & \hspace{-0.8em}$-0.06$ & $5\times 10^{-8}~(d_e)$ \\
$\mu$ & $0.5+0i$ & $1$ & $94.5~(d_n)$ \\
$\tau$ & $0.5+0i$ & 0 & $5.5~(d_n)$ 
 \\ \hline\hline
\end{tabular*}}
\caption{Limits on the vLQ couplings for $2.5$ TeV $U_1$ and $V_2$ from the APV measurement of the $\prescript{133}{55}{\mathbf{Cs}}$ nucleus.
\label{tab:apvtable}}
{\renewcommand\baselinestretch{1.4}\selectfont
\begin{tabular*}{\columnwidth}{l @{\extracolsep{\fill}}cccc}
\hline\hline
 & Couplings & $U_1$  & Couplings& $V_{2}$ \\
\hline
\multirow{2}{*}{$1\sigma$} & $x_{11}^{LL}$                            &    \xmark             & $x_{11}^{LR}$        & \xmark                \\  
                           & $x_{11}^{RR}$                            & $[0.21,0.67]$ & $x_{11}^{RL}$        & $[0.21,0.67]$ \\ \hline
\multirow{2}{*}{$2\sigma$} & $x_{11}^{LL}$                            & $[0,0.40]$    & $x_{11}^{LR}$        & $[0,0.30]$    \\  
                           & $x_{11}^{RR}$                            & $[0,0.78]$    & $x_{11}^{RL}$        & $[0,0.80]$    \\ \hline\hline
\end{tabular*}}
\end{table}

\section{Electric Dipole Moment}\label{sec:EDM}
\noindent
So far, we have assumed all couplings to be real. With this assumption, vLQs contribute to the lepton AMDMs but not their EDMs [as we set $\widetilde \kp_\al=0$, see Eq.~\eqref{eq:SinAMDM}]. If we relax this restriction and allow the couplings to have imaginary parts, the main chirality-flipping term in $d_\ell$, which is proportional to ${\rm Im}(x^{L\al}_{i\ell}{x^{R\overline\al}_{i\ell}}^*)$, starts contributing. To give some idea about the EDM limits, we consider a benchmark set of values for $x_{3\ell}^{LL}$ and $\mathrm{Re}(x_{3\ell}^{RR})$ consistent with the $\Dl a_\ell$ measurements and the LHC bounds for a $2.5$~TeV $U_1$ in Table~\ref{tab:EDM}. The last column shows that the current experimental value of $d_e$ forces the imaginary part of the $x_{31}^{RR}$ coupling to be tiny. However, similar constraints from the $d_\mu$ and $d_\tau$ measurements are weak---they go beyond the perturbative range. Instead, we show the bounds from the neutron EDM ($d_n$) measurements \cite{Abel:2020pzs}, which, as shown in Ref.~\cite{Altmannshofer:2020ywf}, are relatively tighter in these cases but, nevertheless, not very restrictive. The dominant $U_1$ contribution to $d_n$ can be expressed as~\cite{Altmannshofer:2020ywf}
\begin{align}
d_{n}^{\ell} \approx&\ -\frac{g_s^{3}v^2}{(16\pi^2)^2 {m_b}M_{U_1}^2} \beta_n^{\tilde{G}}\sum_\ell m_\ell \, \text{Im}\lt(x^{LX}_{3\ell} x^{RY*}_{3\ell}\rt),
\label{eq:neutronEDM}
\end{align}
where $ \beta^{\tilde{G}}_{n} \approx [0.2, 40] \times 10^{-20} \, \text{e cm}$ \cite{Engel:2013lsa} is the nucleon matrix element. For $V_2$ and other vLQs, the limits are similar. Hence, we do not discuss EDM limits further (interested readers can see Ref.~\cite{Iguro:2023rom}). 

\section{Atomic parity violation}\label{sec:apv}
\noindent
The APV measurement in the $\prescript{133}{55}{\mathbf{Cs}}$ nucleus constrains the vLQ interactions with the first-generation fermions. The tree-level vLQ contribution can be parameterised as follows~\cite{Dorsner:2016wpm}:
\begin{align}
\label{eq:}
\mathcal{L}=\frac{G_{\rm F}}{\sqrt{2}}\overline{e}\gamma^{\mu}\gamma^{5}e\lt(\delta C_{1u}~\overline{u}\gamma_{\mu}u+\delta C_{1d}~\overline{d}\gamma_{\mu}d\rt),
\end{align}
where $\delta C_{1u}$ and $\delta C_{1d}$ come from vLQs. The $U_1$  only contributes to $\delta C_{1d}$ as it cannot simultaneously couple to the up quark and the electron. The $\delta C_{1d}$ from $U_1$ is given as
\begin{equation}
\delta C^{U_1}_{1d}=\frac{v^2}{2M_{U_1}^2}|x^{LL}_{11}|^2-\frac{v^2}{2M_{U_1}^2}|x^{RR}_{11}|^2,   
\end{equation}
where $v$ is the Higgs vacuum expectation value. For $V_2$, $\delta C_{1u}$
and $\delta C_{1d}$ take the following forms:
\begin{align}
\delta C^{V_2}_{1u}&=\frac{v^2}{2M_{V_2}^2}|x^{LR}_{11}|^2,\nn\\
\delta C^{V_2}_{1d}&=\frac{v^2}{2M_{V_2}^2}|x^{LR}_{11}|^2-\frac{v^2}{2M_{V_2}^2}|x^{RL}_{11}|^2.
\end{align}
The above coefficients modify the weak charge for the $\prescript{133}{55}{\mathbf{Cs}}$ nucleus as:
\begin{align}
\label{tab:apvlimit}
\delta Q_W^{\textrm{Cs}}=-2(188~\delta C_{1u}+211~\delta C_{1d})
\end{align}
From experiments, we have 
$Q_W^{\textrm{Cs}}(\textrm{exp})=-72.82 \pm 0.42$~\cite{ParticleDataGroup:2022pth} and $Q_W^{\textrm{Cs}}(\textrm{SM})=-73.33$, where the SM values are, $C_{1u}^{\textrm{SM}}=-0.1887$ and $C_{1d}^{\textrm{SM}}=0.3419$~\cite{Qweak:2018tjf}. In Table \ref{tab:apvtable}, we summarise the limits on the vLQ couplings obtained from the APV measurements for $M_{\ell_q}=2.5$~TeV. The entire $1\sg$ range of the $\delta Q_W^{\textrm{Cs}}$ is positive, $[0.09,0.93]$ whereas the $2\sg$ range stretches from negative to positive values, $[-0.33,1.35]$. This is why we can rule out $x_{11}^{LL}$ (for $U_1$) and $x_{11}^{LR}$ (for $V_2$) single coupling scenarios for the $1\sg$ range of $\delta Q_W^{\textrm{Cs}}$.

\begin{table*}
\caption{\label{tab:feynrules} 
Illustrative Feynman rules and momentum conventions.}
\centering{\footnotesize\renewcommand\baselinestretch{2}\selectfont
\begin{tabular*}{0.75\textwidth}{@{\extracolsep{\fill}}c c  c c }
\hline \hline 
Diagram & Rule & Diagram & Rule \\
\hline
\begin{tikzpicture}[baseline={(current bounding box.center)}]
    \begin{feynman}
    \vertex (a) ;
    \vertex [above  left=of a] (b) {\(e_L\)};
    \vertex [below  left=of a] (d) {\(\overline{q_L}\)};
    \vertex [      right=of a] (f) ;
    \diagram* {
        (b) -- [fermion,momentum=\(p_{1}\)] (a) -- [fermion,reversed momentum=\(p_{2}\)] (d),
        (a) -- [boson, edge label'=\(V_{\ell_q}\),reversed momentum=\(p_3\),purple] (f),
        
    };
    \end{feynman}
    \end{tikzpicture} & $x^{LL}\Gamma_\mu$ & \begin{tikzpicture}[baseline={(current bounding box.center)}]
    \begin{feynman}
    \vertex (a) ;
    \vertex [above  left=of a] (b) {\(e_R\)};
    \vertex [below  left=of a] (d) {\(\overline{q_R}\)};
    \vertex [      right=of a] (f) ;
    
    \diagram* {
        (b) -- [fermion,momentum=\(p_{1}\)] (a) -- [fermion,reversed momentum=\(p_{2}\)] (d),
        (a) -- [boson, edge label'=\(V_{\ell_q}\),reversed momentum=\(p_3\),purple] (f),
        
    };
    \end{feynman}
    \end{tikzpicture}& $x^{RR}\Gamma_\mu$\\
\begin{tikzpicture}[baseline={(current bounding box.center)}]
    \begin{feynman}
    \vertex (a) ;
    \vertex [above  left=of a] (b) {\(e_L\)};
    \vertex [below  left=of a] (d) {\(\overline{q^c_R}\)};
    \vertex [      right=of a] (f) ;
    
    \diagram* {
        (b) -- [fermion,momentum=\(p_{1}\)] (a) -- [fermion,reversed momentum=\(p_{2}\)] (d),
        (a) -- [boson, edge label'=\(V_{\ell_q}\),reversed momentum=\(p_3\),purple] (f),
        
    };
    \end{feynman}
    \end{tikzpicture}& $x^{LR}\Gamma_\mu$ & \begin{tikzpicture}[baseline={(current bounding box.center)}]
    \begin{feynman}
    \vertex (a) ;
    \vertex [above  left=of a] (b) {\(e_R\)};
    \vertex [below  left=of a] (d) {\(\overline{q^c_L}\)};
    \vertex [      right=of a] (f) ;
    
    \diagram* {
        (b) -- [fermion,momentum=\(p_{1}\)] (a) -- [fermion,reversed momentum=\(p_{2}\)] (d),
        (a) -- [boson, edge label'=\(V_{\ell_q}\),reversed momentum=\(p_3\),purple] (f),
        
    };
    \end{feynman}
    \end{tikzpicture}& $x^{RL}\Gamma_\mu$\\
        \begin{tikzpicture}[baseline={(current bounding box.center)}]
    \begin{feynman}
    \vertex (a) ;
    \vertex [above  left=of a] (b) {\(q\)};
    \vertex [below  left=of a] (d) {\(\overline{q}\)};
    \vertex [      right=of a] (f) ;
    \diagram* {
        (b) -- [fermion,momentum=\(p_{1}\)] (a) -- [fermion,reversed momentum=\(p_{2}\)] (d),
        (a) -- [photon, edge label'=\(\gamma_\mu\),reversed momentum=\(p_3\)] (f),
        
    };
    \end{feynman}
    \end{tikzpicture} & $\gamma_{\mu}$ & \begin{tikzpicture}[baseline={(current bounding box.center)}]
    \begin{feynman}
    \vertex (a) ;
    \vertex [above  left=of a] (b) {\(q^c\)};
    \vertex [below  left=of a] (d) {\(\overline{q^c}\)};
    \vertex [      right=of a] (f) ;    
    \diagram* {
        (b) -- [fermion,momentum=\(p_{1}\)] (a) -- [fermion,reversed momentum=\(p_{2}\)] (d),
        (a) -- [photon, edge label'=\(\gamma_\mu\),reversed momentum=\(p_3\)] (f),
        
    };  
    \end{feynman}
    \end{tikzpicture}& $-\gamma_{\mu}$\\  
\begin{tikzpicture}[baseline={(current bounding box.center)}]
    \begin{feynman}
    \vertex (a) ;
    \vertex [above  left=of a] (b) {\(V_{\ell_q}\)};
    \vertex [below  left=of a] (d) {\(V_{\ell_q}\)};
    \vertex [      right=of a] (f) ;
    
    \diagram* {
        (b) -- [boson,momentum=\(p_{1}\),purple] (a) -- [boson,reversed momentum=\(p_{2}\),purple] (d),
        (a) -- [photon, edge label'=\(\gamma_\mu\),reversed momentum=\(p_3\)] (f),
        
    };
    \end{feynman}
    \end{tikzpicture} & \multicolumn{3}{c}{${\mathcal{P}}_{\mu,\alpha,\beta}=(\kappa_Y p_{3\beta}-p_{1\beta})g_{\mu\alpha}-(\kappa_Y p_{3\alpha}-p_{2\alpha})g_{\mu\beta} +(p_{1\mu}-p_{2\mu})g_{\beta\alpha}+\widetilde{\kappa_y}\epsilon_{\mu\alpha\beta\delta}p_3^{\delta}$}\vspace{-0.5cm}\\  
 \multirow{3}{*}{\begin{tikzpicture}[baseline={(current bounding box.center)}]
    \begin{feynman}
    \vertex (a) {\(V_{\ell_q}\)};
    \vertex [      right=of a] (f){\(V_{\ell_q}\)} ;
    
    \diagram* {
        (a) -- [boson,momentum=\(p\),purple] (f),
        
    };
    \end{feynman}
    \end{tikzpicture}} & \multirow{3}{*}{$V^{\mu\nu}(p)=\dfrac{1}{p^2-M_V^2}\lt(g^{\mu\nu}-\dfrac{p^\mu p^\nu}{M_V^2}\rt)$}& \begin{tikzpicture}[baseline={(current bounding box.center)}]
    \begin{feynman}
    \vertex (a) {\(\overline{q}\)};
    \vertex [      right=of a] (f){\(q\)} ;
    
    \diagram* {
        (a) -- [fermion,momentum=\(p\)] (f),
        
    };
    \end{feynman}
    \end{tikzpicture} & $S(p)=\dfrac{i(\slashed{p}+m)}{p^{2}-m^{2}}$\\&&&\\    & & \begin{tikzpicture}[baseline={(current bounding box.center)}]
    \begin{feynman}
    \vertex (a) {\(\overline{q^c}\)};
    \vertex [      right=of a] (f){\(q^c\)} ;
    
    \diagram* {
        (a) -- [fermion,momentum=\(p\)] (f),        
    };
    \end{feynman}
    \end{tikzpicture} & $S(p)=\dfrac{i(-\slashed{p}+m)}{p^{2}-m^{2}}$\\\hline\hline
\end{tabular*}}
\end{table*}

\begin{figure}[!t]
\centering
\captionsetup[subfigure]{labelformat=empty}
\hspace{-0.5cm}\includegraphics[width=0.7\columnwidth]{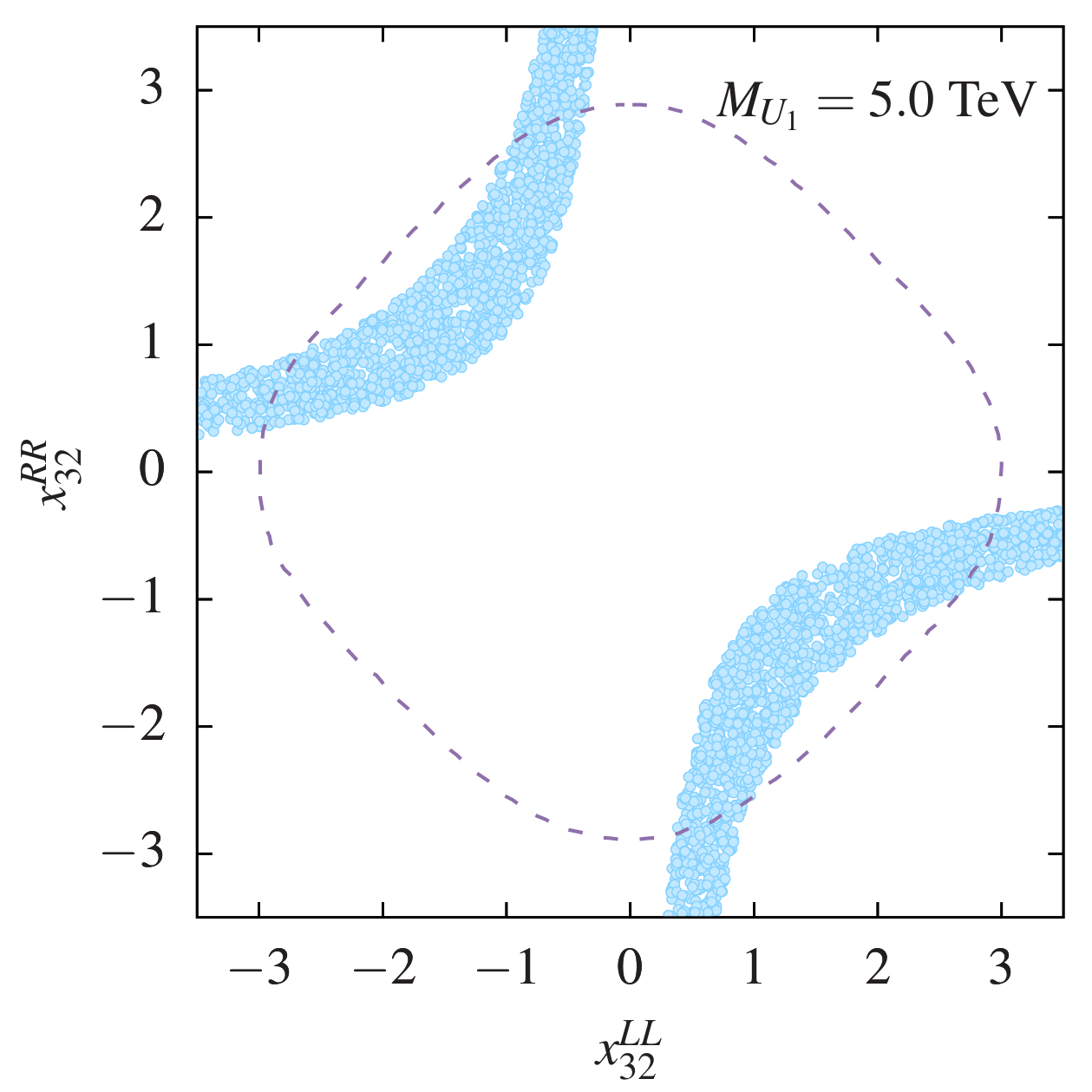}
\caption{\label{fig:gM2a5000} Case of heavy vLQs: an illustration with $5$ TeV $U_1$---the $(g-2)_\m$ parameter space and the corresponding indirect LHC bound. For heavy vLQs, the indirect bounds are more restrictive than the direct ones. Compared to Fig.~\ref{fig:gM2}, the necessary couplings have increased, but the LHC limit has also loosened.}
\end{figure}
\section{Summary and Conclusions}\label{sec:conclu}
\noindent
In this paper, we examined the contributions of vLQs to the dipole moments of charged leptons---particularly $e$ and $\m$, which are measured precisely---in light of the high-energy dilepton data from the LHC. We derived the contributions of weak-singlet, doublet, and triplet vLQs to $(g-2)_\ell$ in the presence of additional gauge couplings $\kappa$ and $\widetilde\kappa$. The parameter space favoured by the $(g-2)_\ell$ measurements is constrained by the LHC dilepton bounds. Following the method outlined in Ref.~\cite{Bhaskar:2023ftn}, we considered both the dilepton-dijet and the dilepton resonance search data to derive the LHC bounds. In the parameter range of our interest, the LHC limits dominantly come from the $t$-channel vLQ-exchange diagrams and their interference with the SM background, making the LHC constraints insensitive to $\kp$ and $\widetilde\kappa$. We set $\kp=1$ and $\widetilde{\kp}=0$ throughout our analysis, as the one-loop vLQ contributions to leptonic dipole moments remain independent of the loop-momentum cut-off in this limit. In this limit, our results also agree with Ref.~\cite{Altmannshofer:2020ywf}. 

In single-coupling scenarios, $(g-2)_\ell$ gets contributions from chirality-preserving terms that are suppressed by $m_\ell^2/M_{\ell_q}^2$. Consequently, no single-coupling scenario can account for the measured $(g-2)_\ell$ values for perturbative couplings. However, vLQs that couple to both left- and right-handed leptons can resolve the $(g-2)_\mu$ anomaly through chirality-flipping contributions in principle. Of the five possible vLQs, this is only possible for $U_1$ and $V_2$. These two vLQs can explain the observed positive shift in $\Dl a_\mu$ with perturbative couplings in two-coupling scenarios, provided they interact with third-generation quarks, as is the case in the down-aligned $x_{32}^{LX}$, $x_{32}^{RY}$ scenarios ($\{X,R\}=\{L,R\}$ for $U_1$ and $\{X,Y\}=\{R,L\}$ for $V_2$). One could, in principle, consider similar up-aligned scenarios as well. However, those will contribute significantly to a range of observables in meson mixings ($B_0$-$\overline{B}_0$, $B_s$-$\overline{B}s$, $K_0$-$\overline{K}_0$, etc.) and decays ($R_{K^{(*)}}$, $R^\nu_{K^{(*)}}$, $B_s\to \mu\mu$, etc.). (The down-aligned scenarios also contribute to $D_0$-$\overline{D}_0$ mixing and decays such as $D_0\to \mu\mu$ and $B\to \mu\nu$, but off-diagonal Cabibbo-Kobayashi-Maskawa matrix elements strongly suppress the contributions.) The $Z$-boson decays also constrain these scenarios~(see, e.g., Ref.~\cite{Bansal:2018nwp} for a similar case with scalar LQs) but the dominant contributions from the top-quark loops are absent in the down-aligned scenarios. While it is certainly interesting to investigate the role of various low-energy bounds in different scenarios, a comprehensive analysis lies beyond the scope of this paper.

The AMDM measurements for electrons show a mild discrepancy, with conflicting results between the $\prescript{133}{55}{\mathbf{Cs}}$-based and $\prescript{87}{37}{\mathbf{Rb}}$-based experiments. We found that two-coupling (chirality-flipping) scenarios can account for the measured values of $(g-2)_e$, yielding perturbative coupling solutions, especially when the vLQs couple to either second- or third-generation quarks. We performed parameter scans for $2.5$ TeV $U_1$ and $V_2$ to show how these vLQs can accommodate the precisely measured AMDMs of the electron and the muon without conflicting with the LHC bounds. The LHC bounds prevent these vLQs from being much lighter than $\sim (2-2.5)$ TeV. On the other hand, since heavier vLQs require large couplings to accommodate the $(g-2)_\ell$ measurements (see Fig.~\ref{fig:gM2a5000}, which shows the trend), they cannot be arbitrarily heavy if we demand the couplings remain in the perturbative domain. For example, for $x_{32}^{LL}x_{32}^{RR}< \sqrt{16\pi^2}$, the $U_1$ cannot be heavier than about $13$ TeV. Similarly, for $x_{32}^{LR}x_{32}^{RL}< \sqrt{16\pi^2}$, $M_{V_2} \lesssim 17$ TeV. However, such heavy vLQs will be much beyond the direct reach of the current collider experiments. 

If we assume the new couplings are complex in general, vLQs contribute to lepton EDMs. However, except for the electron, the current lepton EDM measurements are too weak to restrict the $\m$ and $\tau$ couplings within the perturbative range (which are ruled out by the LHC data anyway). Only the electron EDM measurement can put a tight bound on the imaginary parts of the electron couplings if we assume $\widetilde \kp=0$. The electron couplings with the first-generation quarks are also constrained to be within the perturbative range by the APV measurements.

\section*{acknowledgement}
\noindent We thank Arijit Das for helping us with the direct LHC limits. D.D. would like to thank the SERB/ANRF, Govt. of India for the SRG Grant Order No. SRG/2023/001318 and IIIT Hyderabad for the Seed Grant No. IIIT/R\&D Office/Seed-Grant/2021-22/013. T.M. is supported by the intramural grant from IISER-TVM.   
\appendix
\section{Anomalous magnetic-dipole moment of leptons in vLQ scenarios}\label{sec:generalgM2}
\noindent
The general lagrangian for a $SU(2)_L$-singlet, -doublet, or -triplet vLQ interaction with the SM quarks and charged leptons can be written as 
\begin{align}
\mathcal{L} \supset&\ x_{ij}^{LL} \overline{q}_{L}^i\gamma_{\mu}\ell_{L}^jV^{\mu}_0+x_{ij}^{RR} \overline{q}_{R}^i\gamma_{\mu}\ell_{R}^jV^{\mu}_0\nn\\
&\ +x_{ij}^{RL} {\overline{q}^{c}}^{i}_{R}\gamma_\mu\ell_{L}^jV^{\mu}_2+x_{ij}^{LR} {\overline{q}^{c}}^i_{L}\gamma_\mu\ell_{R}^jV^{\mu}_2+ h.c.
\end{align}
With this, the generic vLQ contribution to $a_\ell$ at the leading order can be written as
\begin{align}
    a_{\ell}=& \frac{N_{c}}{16\pi^{2}}\mathlarger{\sum}_{i=1}^3\mathlarger{\sum}_{\al}^{L,\,R}\Bigg[2{\rm Re} (x^{L\al}_{i\ell}{x^{R\overline\alpha}_{i\ell}}^*)\frac{m_{\ell}m_{q_{i}}}{M_{\ell_q}^{2}}\Big\{2Q_{q}\nonumber\\
    &\ +Q_{\ell_q}\left((1-\kappa_{\al})\ln\left(\frac{\Lambda^{2}}{M_{\ell_q}^{2}}\right)+\frac{1-5\kappa_{\al}}{2}\right)\Big\} \nonumber\\
    &\ +(1-2\delta_{L\alpha})(|x^{L\al}_{i\ell}|^2+|x^{R\overline\alpha}_{i\ell}|^2)\frac{m_{\ell}^{2}}{M_{\ell_q}^{2}}\Bigg\{\frac{4}{3}Q_{q}\nonumber\\
    &\ +Q_{\ell_q}\left((1-\kappa_{\al})\ln\left(\frac{\Lambda^{2}}{M_{\ell_q}^{2}}\right)-\frac{1+9\kappa_{\al}}{6}\right)\Bigg\}\nonumber\\
    &+  2\widetilde\kappa_\alpha\, Q_{\ell_q} {\rm Im}(x^{L\al}_{i\ell}{x^{R\overline\alpha}_{i\ell}}^*)\frac{m_{\ell}m_{q_{i}}}{M_{\ell_q}^{2}}\left(\ln{\frac{\Lambda^{2}}{M_{\ell_q}^{2}}}-\frac{1}{2}\right)\Bigg],\label{eq:LQAMDM}
\end{align}
where $\overline \alpha$ stands for the opposite chirality to $\alpha$, $\kappa_{\alpha} = \kappa_Y$ and $\widetilde\kappa_{\alpha} = \widetilde\kappa_Y$ if $\alpha = L$ (e.g., for singlet vLQs), and $ \kappa_{\alpha} =\kappa_{eff}$ and $\widetilde\kappa_{\alpha} = \widetilde\kappa_{eff}$ otherwise (i.e., for doublet vLQs). Similarly, the contribution to $d_\ell$ can be written as
\begin{align}
d_{\ell}=&\ \frac{eN_{c}}{16\pi^{2}}\mathlarger{\sum}_{i=1}^3\mathlarger{\sum}_{\al}^{L,\,R}\Bigg[{\rm Im}(x^{L\al}_{i\ell}{x^{R\overline\al}_{i\ell}}^*)\frac{m_{q_{i}}}{M_{\ell_q}^{2}}\Bigg\{2Q_{q}\nonumber\\ 
&\ +Q_{\ell_q}\left((1-\kappa_\al)\ln\left(\frac{\Lambda^{2}}{M_{\ell_q}^{2}}\right)+\frac{1-5\kappa_{\al}}{2}\right)\Bigg\} \nonumber \\
&+(2\delta_{L\alpha}-1)Q_{\ell_q}\widetilde{\kappa}_\al(|x^{L\al}_{i\ell}|^2+|x^{R\overline\al}_{i\ell}|^2)\frac{m_{\ell}}{M_{\ell_q}^{2}}\left(\frac{1}{2}\ln{\frac{\Lambda^2}{M_{\ell_q}^2}}-\frac{3}{4}\right)\nonumber\\
&-  \widetilde \kappa_\al Q_{\ell_q} {\rm Re}(x^{L\al}_{i\ell}{x^{R\overline\al}_{i\ell}}^*)\frac{m_{q_{i}}}{M_{\ell_q}^{2}}\left(\ln{\frac{\Lambda^{2}}{M_{\ell_q}^{2}}}-\frac{1}{2}\right)\Bigg].\label{eq:SinAMDM}
\end{align}
We evaluate the loop contributions following the convention in Table~\ref{tab:feynrules} and using~\textsc{Package-X}~\cite{Patel:2016fam} in the limit $m_\m, m_q \ll M_{\ell_q}$, where $m_q$ and $M_{\ell_q}$ are the masses of the quark and the vLQ in the loop.

\bibliography{reference}

\end{document}